%% file: TNG_PMs.tex
\documentclass[fleqn,usenatbib]{mnras}

\usepackage{newtxtext,newtxmath}

\usepackage[T1]{fontenc}
\usepackage{ae,aecompl}

\usepackage{graphicx}	
\usepackage{amsmath}	
\usepackage{amssymb}	

\usepackage[dvipsnames]{xcolor}

\usepackage{tabularx}
\usepackage{nicefrac}

\title[IllustrisTNG post-merger galaxies]{Interacting galaxies in the IllustrisTNG simulations -- II: Star formation in the post-merger stage.}

\author[M. H. Hani et al.]{
  \parbox[t]{\textwidth}{
  {Maan H. Hani,$^{1}$\thanks{E-mail: mhani@uvic.ca}\thanks{Vanier Scholar}}
  {Hayman Gosain,$^{1,2}$ }
  {Sara L. Ellison,$^{1}$ }
  {David R. Patton,$^{3}$ }\\
  {Paul Torrey$^{4}$ }
  }
\\\\
$^{1}$Department of Physics and Astronomy, University of Victoria, Victoria, British Columbia, V8P 1A1, Canada\\
$^{2}$Department of Physical Sciences, Indian Institute of Science Education and Research Mohali, Manauli, Punjab 140306, India\\
$^{3}$Department of Physics and Astronomy, Trent University, 1600 West Bank Drive, Peterborough, ON K9L 0G2, Canada\\
$^4$Department of Astronomy, University of Florida, 211 Bryant Space Sciences Center, Gainesville, FL, USA\\
}

\date{Accepted XXX. Received YYY; in original form ZZZ}

\pubyear{2020}

\begin{document}
\label{firstpage}
\pagerange{\pageref{firstpage}--\pageref{lastpage}}
\maketitle



\input{./abstract.tex}
\input{./introduction.tex}
\input{./methods.tex}

\input{./results.tex}

\input{./discussion.tex}
\input{./conclusions.tex}
\input{./acknowledgements.tex}



\bibliographystyle{mnras}
\bibliography{references}



\bsp	
\label{lastpage}
\end{document}

%% file: abstract.tex
\begin{abstract}
Galaxy mergers are a major evolutionary transformation whose effects are borne out by a plethora of observations and numerical simulations. However, most previous simulations have used idealised, isolated, binary mergers and there has not been significant progress on studying statistical samples of galaxy mergers in large cosmological simulations. We present a sample of 27,691 post-merger (PM) galaxies ($0\le z \le 1$) identified from IllustrisTNG: a cosmological, large box, magneto-hydrodynamical simulation suite. The PM sample spans a wide range of merger and galaxy properties ($M_\star$, $\mu$, $f_\mathrm{gas}$). We demonstrate that star forming (SF) PMs exhibit enhanced star formation rates (SFRs) on average by a factor of $\sim 2$, while the passive PMs show no statistical enhancement. We find that the SFR enhancements: (1) show no dependence on redshift, (2) anti-correlate with the PM's stellar mass, and (3) correlate with the gas fraction of the PM's progenitors. However, SF PMs show stronger enhancements which may indicate other processes being at play (e.g., gas phase, feedback efficiency). Although the SFR enhancement correlates mildly with the merger mass ratio, the more abundant minor mergers ($0.1 \le \mu < 0.3 $) still contribute $\sim 50\%$ of the total SFR enhancement. By tracing the PM sample forward in time, we find that galaxy mergers can drive significant SFR enhancements which decay over $\sim 0.5$ Gyr independent of the merger mass ratio, although the decay timescale is dependent on the simulation resolution. The strongest merger-driven starburst galaxies evolve to be passive/quenched on faster timescales than their controls.
\end{abstract}


%% file: introduction.tex
\section{Introduction}
\label{sec:intro}
\noindent

Galaxy mergers are at the foundation of the hierarchical model for structure formation \citep{1978MNRAS.183..341W, 1993MNRAS.262..627L}. Through mergers, galaxies can grow their stellar, gas, and dark matter content. However, the effects of galaxy mergers are not limited to the growth of galaxies. There exists a rich body of theoretical predictions and observational evidence linking mergers to triggering and accelerating gas evolution (i.e., depletion, ejection, enrichment, cooling/heating) thus confirming that galaxy mergers are a major rite of passage in galaxy evolution.

Observationally, a profusion of studies report several well accepted observational signatures of galaxy mergers. A dilution in the central metallicity has been reported by observational studies of galaxy pairs \citep[e.g.,][]{2006AJ....131.2004K, 2008AJ....135.1877E, 2012MNRAS.426..549S, 2018MNRAS.478.3447E}, and post-mergers \citep[][]{2013MNRAS.435.3627E, 2019MNRAS.482L..55T}. Additionally, merging galaxies exhibit enhanced morphological disturbances \citep[][]{2014MNRAS.445.1157C, 2016MNRAS.461.2589P}, increased atomic and molecular gas fractions \citep[e.g.,][]{2015MNRAS.448..221E, 2018MNRAS.478.3447E, 2018ApJ...868..132P, 2018MNRAS.476.2591V, 2018MNRAS.480..947D, 2019MNRAS.489.1099D}, elevated AGN fractions \citep[e.g.,][]{2011MNRAS.418.2043E, 2014MNRAS.441.1297S, 2019MNRAS.487.2491E}, and enhanced SFRs \citep[e.g.,][]{2008AJ....135.1877E, 2013MNRAS.435.3627E, 2013MNRAS.433L..59P,  2015MNRAS.454.1742K, 2019MNRAS.482L..55T} when compared to their non-interacting counterparts. Coupled with enhanced star formation and AGN activity, feedback processes (e.g., stellar \& AGN feedback) play a vital role in the evolution of mergers. Galactic outflows have been linked to enhanced star formation \citep[e.g.,][]{2005ApJ...621..227M,2005ApJS..160..115R, 2009ApJ...697.2030S} and AGN activity \citep[e.g.,][]{2005ApJ...632..751R, 2013ApJ...776...27V, 2017ApJ...839..120W}.

The observational signatures of galaxy interactions and mergers are complemented by a well-posed theoretical framework. Galaxy interactions trigger strong non-axisymmetric features which create tidal torques capable of driving instabilities in the dynamically cool inter-stellar medium (ISM) thus causing strong inflows of gas into the galactic centre \citep[e.g.,][]{1989Natur.340..687H, 1991ApJ...370L..65B,1996ApJ...464..641M, 2018MNRAS.479.3952B}. The gravitational instability-driven gas inflow manifests in simulations as a dilution in central metallicities \citep[e.g.,][]{2010A&A...518A..56M, 2012ApJ...746..108T, 2015MNRAS.448.1107M, 2018MNRAS.479.3381B}, an enhancement in central star formation rates \citep[SFRs; e.g.,][]{2008MNRAS.384..386C, 2008A&A...492...31D, 2015MNRAS.448.1107M, 2016MNRAS.462.2418S}, and an increase in accretion onto the central super-massive black hole thus giving rise to an active galactic nucleus \citep[AGN; e.g.,][]{2005Natur.433..604D, 2010MNRAS.407.1529H}. The associated feedback processes often manifest (or are implemented) as outflows \citep[e.g.,][]{2017MNRAS.465.1682H, 2019MNRAS.485.1320M} which can shape the ISM and even the circum-galactic medium \citep[e.g.,][]{2004ApJ...607L..87C, 2018MNRAS.475.1160H}. Additionally, mergers' descendants possess unique morphological disturbances such as shells, ripples, tidal tails, and tidal plumes \citep[e.g.,][]{2007A&A...468...61D, 2008MNRAS.391.1137L, 2010MNRAS.404..575L, 2010MNRAS.404..590L, 2018MNRAS.480.1715P}.

Numerical simulations provide a controlled environment to experiment and investigate the physical process at play during a galaxy merger. Such controlled experiments are often carried out in simulations of binary mergers where the same galaxies that partake in the merger are evolved in isolation \citep[e.g.,][]{2012ApJ...746..108T, 2013MNRAS.433L..59P, 2015MNRAS.448.1107M, 2019MNRAS.485.1320M}. While isolated binary merger suites provide a controlled environment to study the details of galaxy mergers, they are limited in their level of realism (non-cosmological environment, lack of circum-galactic gas and fully realistic range of parameters such as galaxy morphology, gas fraction, and orbital geometry), and therefore do not represent the majority of observed galaxies \citep[see][]{2013MNRAS.436.1765M}. 

Cosmological zoom-in simulations provide a more realistic, yet still limited (e.g., galaxy morphology, gas fraction, orbital geometry), approach to study galaxy mergers where galaxies are evolved from cosmological initial conditions \citep[e.g.,][]{2016MNRAS.462.2418S, 2018MNRAS.479.3381B, 2018MNRAS.475.1160H}. One particular drawback of such simulations is the limited ability to trace the galaxies for a long time after the merger. In cosmological zoom-in simulations, the effects of the merger are often compared to a pre-merger state \citep[e.g.,][]{2016MNRAS.462.2418S, 2018MNRAS.475.1160H}; however, on long timescales, post-mergers evolve significantly and therefore should not be compared to the pre-merger state.

While both binary merger simulations and cosmological zoom-in simulations provide powerful tools to study galaxy mergers at high resolution (spatial and temporal), they both lack the large and diverse galaxy samples of large-box cosmological simulations. Recent studies have begun to investigate the effect of galaxy mergers in large cosmological boxes \citep[e.g.,][]{2019MNRAS.tmp.3093B, TNGpairs, 2019MNRAS.490.2139R}.

The work presented here leverages the sizeable sample of galaxy mergers available in large cosmological box hydrodynamical simulations to investigate the impact of mergers on galactic star formation during the post-merger stage (i.e., post-coalescence). Using numerical simulations mitigates the unconstrained merger timescales in observational studies, while the large simulation box provides a diverse and representative galaxy sample therefore alleviating a major challenge for idealised and cosmological zoom-in simulations. We identify galaxy mergers in the IllustrisTNG simulation suite \citep{2018MNRAS.480.5113M, 2018MNRAS.477.1206N, 2018MNRAS.475..624N, 2018MNRAS.475..648P, 2018MNRAS.475..676S, 2019ComAC...6....2N} and then analyse our post-merger galaxy sample by implementing a methodology akin to observations. By comparing the post-merger star formation rates to controls which have not undergone a merger, we are able to isolate the effects of the merger on star formation rates \citep[e.g.,][]{TNGpairs}. 

The manuscript is structured as follows: We first introduce the methodology in Section \ref{sec:methods}. In Section \ref{sec:results} we present the results highlighting the role of galaxy mergers inducing star formation and we investigate major drivers to the strength of the SFR enhancements. Section \ref{sec:discussion} discusses the effects of control matching and simulation resolution, and ties our results to observational studies. Finally, we summarise our conclusions in Section \ref{sec:conclusions}.

%% file: methods.tex
\section{Methods}
\label{sec:methods}
\noindent
The work presented here investigates the signatures of galaxy mergers during the post-merger stage. We identify galaxy mergers in the IllustrisTNG simulation suite \citep{2018MNRAS.480.5113M, 2018MNRAS.477.1206N, 2018MNRAS.475..624N, 2018MNRAS.475..648P, 2018MNRAS.475..676S, 2019ComAC...6....2N} and compare the properties of the mergers' descendants to galaxies which have not undergone a merger in their recent evolution. In this section, we describe the numerical simulations (i.e., IllustrisTNG simulations), the details of our merger-identification, and the process of generating a statistical control galaxy sample for comparison.

\subsection{Numerical simulations: TNG100-1 \& TNG300-1}
\noindent
The work presented here is primarily focused on quantifying the effects of galaxy mergers beyond coalescence (post-mergers). We employ the IllustrisTNG simulation suite \citep{2018MNRAS.480.5113M, 2018MNRAS.477.1206N, 2018MNRAS.475..624N, 2018MNRAS.475..648P, 2018MNRAS.475..676S, 2019ComAC...6....2N} which consists of several large box cosmological magneto-hydrodynamical simulations. We focus our analysis on the highest resolution runs of the largest two volumes of the IllustrisTNG simulations -- TNG100-1 and TNG300-1. Using such large volumes ($110.7^3$ Mpc$^3$ and $302.6^3$ Mpc$^3$ for TNG100-1 and TNG300-1, respectively) ensures a substantial sample of galaxies ideal for a statistical study of galaxy mergers. The galaxy sample is simulated in a fully cosmological environment, and it has realistic evolutionary histories, and galactic properties \citep[e.g., ][]{2018MNRAS.475..624N, 2018MNRAS.475..648P, 2018MNRAS.475..676S,  2019MNRAS.485.4817D, 2019MNRAS.489.1859H, 2019MNRAS.483.4140R, 2019MNRAS.487.5416T, 2019MNRAS.484.5587T}. While TNG300-1 provides larger galaxy and merger samples, TNG100-1 allows us to investigate the effects of galaxy mergers at higher mass and spatial resolutions. TNG100-1 (TNG300-1) has a dark matter mass resolution $m_\mathrm{dm}=7.5\times10^6$ M$_\odot$ ($m_\mathrm{dm} = 5.9\times10^7$ M$_\odot$), and a baryonic mass resolution $m_\mathrm{b}\sim 10^6$ M$_\odot$ ($m_\mathrm{b}\sim10^7$ M$_\odot$). We also use TNG100-2 to ensure model convergence and investigate resolution effects; TNG100-2 is a re-run of TNG100-1 at a roughly equal resolution to TNG300-1. All the simulations have a temporal resolution $\sim 162$ Myr.

For most of our analysis we will report results from TNG300-1 as our fiducial simulation because the large box size provides exquisite statistics. Nonetheless, the results are broadly consistent between TNG300-1 and TNG100-1. We refer the reader to Section \ref{sec:discussion:restest} where we discuss the effects of varying the simulation resolution on our conclusions.

The IllustrisTNG galaxy formation model which was introduced in  \citet{2017MNRAS.465.3291W} and \citet{2018MNRAS.473.4077P} builds on its predecessor, the Illustris model \citep{2013MNRAS.436.3031V, 2014MNRAS.438.1985T}, with several additions and modifications to the numerical framework and physical model. The IllustrisTNG physical model includes:
\begin{enumerate}
    \item{Star formation: Star formation occurs in a pressurised interstellar medium (ISM) for $n_\mathrm{H} \gtrsim 0.1$ cm$^{-3}$ following the \citet{2003MNRAS.339..289S} formalism. Star particles represent stellar populations with a Chabrier \citep{2003PASP..115..763C} initial mass function.}
    \item{Galactic winds: Stellar feedback is implemented using hydrodynamically decoupled winds which transport mass, momentum, metals, and thermal energy.}
    \item{Metal enrichment: Metals are returned to the ISM by supernovae (SN) type Ia, SN type II, and asymptotic giant branch (AGB) stars.}
    \item{Gas cooling and heating: Gas can cool through primordial channels as well as metal-line cooling. Gas heating is calculated assuming a superposition of a redshift-dependent, spatially uniform UV background \citep{2009ApJ...703.1416F}, and the active galactic nucleus (AGN) radiation field \citep{2013MNRAS.436.3031V}.}
    \item{Growth and feedback of black holes (BHs): Accretion onto BH sink particles is described by Bondi-Hoyle accretion. The associated feedback from AGN employs kinetic feedback (low accretion rates), and thermal feedback (high accretion rates).}
\end{enumerate}

For a detailed description of the simulations, and physical model, we refer the reader to the IllustrisTNG introduction and methods papers: \citet{2017MNRAS.465.3291W, 2018MNRAS.480.5113M, 2018MNRAS.477.1206N, 2018MNRAS.475..624N, 2018MNRAS.473.4077P, 2018MNRAS.475..648P, 2018MNRAS.475..676S}, and references therein.

\subsection{Merger identification}
\label{sec:methods:mergers}
\noindent
We identify galaxy-galaxy mergers in IllustrisTNG from the publicly available merger trees created using \textsc{Sublink} \citep{2015MNRAS.449...49R}. The \textsc{Sublink} merger trees link sub-haloes to their progenitors and descendants. Consequently, galaxy mergers are defined as nodes within a tree; viz. a merger occurs when a sub-halo has two distinct direct progenitors. Note that we use ``post-merger'' to refer to galaxies immediately after the merger (i.e., first snapshot after the merger) which corresponds to $\le 162$ Myr post-merger at the snapshot time resolution of IllustrisTNG. In Section \ref{sec:results:evolution}, we follow the evolution of post-mergers and hence, in that section only, ``post-merger'' refers to galaxies which have undergone a merger in their recent past. Therefore, post-mergers (merger remnants) are parametrized by: 
\begin{itemize}
    \item[$\bullet$]{$z$: The redshift of the post-merger. We include in our analysis  post-mergers at $z\le 1$ which allows us to study the redshift evolution (or lack thereof) of mergers and their descendants.}
    \item[$\bullet$]{$M_\star$: The stellar mass of the post-merger. Various studies have noted the spurious assignment of mass caused by sub-haloes in close proximity when using halo finders (i.e., numerical stripping). For example, the dark matter mass of satellites has been shown to correlate with the distance to their host galaxies \citep{2007MNRAS.379.1464S, 2009MNRAS.395.1376W}. Numerical stripping has also been demonstrated to affect stellar mass estimates \citep[i.e.,][]{2015MNRAS.449...49R}. We circumvent the subtleties of stellar mass calculations by adopting the maximum stellar mass over the past $0.5$ Gyr ($M_\star ^\mathrm{max}$) for all mass ratio calculations following \citet{TNGpairs}. This approach is similar to \citet{2015MNRAS.449...49R}, with an additional restriction on lookback time; i.e., we minimise the effects of numerical stripping yet we still account for physical stripping by limiting the lookback time. Driven by the simulation resolution, galaxies with stellar masses $\log(M_\star / \mathrm{M_\odot}) \ge 9$ are reliably resolved. Therefore, we limit our analysis to post-mergers with stellar mass $10^{10} \le M_\star / \mathrm{M}_\odot \le 10^{12}$ and progenitors with stellar mass $M_\star  \ge 10^{10} \mathrm{M}_\odot $ (see the mass ratio description).}
    \item[$\bullet$]{$\mu$: The stellar mass ratio of the progenitors. Note that we define the mass ratio to be $0 < \mu \le 1$. In cases where more than two direct progenitors are found, we parametrize the merger remnant using only the most major merger. The masses used to calculate $\mu$ correspond to the stellar mass within twice the stellar half-mass radius. Our analysis includes descendants of mergers with $\mu\ge 0.1$ which is a requisite for investigating the varying effects of mass ratio in galaxy mergers. The simulation's mass resolution limit, coupled with the mass ratio range determines the stellar mass range of the post-merger sample and their progenitors: $10^{10} \le M_\star / \mathrm{M}_\odot \le 10^{12}$.}
\end{itemize}

We are primarily interested in studying the effects of galaxy mergers on star formation during the post-merger stage, therefore we exclude from our sample post-mergers that are currently undergoing new close interactions but have not yet fully merged. Particularly, we ignore post-mergers that are overlapping with another galaxy which hinders our ability to separate the effects of the current interaction from those of the merger. Following the methodology of \citet{TNGpairs}, we define:
\begin{equation}
    r_\mathrm{sep} = \frac{r}{R^\mathrm{host}_{1/2} + R^\mathrm{comp}_{1/2}}
\end{equation}
where $r$ is the separation of the host (post-merger) from its nearest neighbour (companion), $R^\mathrm{host}_{1/2}$ is the stellar half-mass radius of the host (post-merger), and $R^\mathrm{comp}_{1/2}$ is the stellar half-mass radius of the nearest neighbour. We exclude post-mergers with $r_\mathrm{sep}\le 2$ from our sample. In addition, we ignore galaxies with unresolved SFRs, viz. $SFR < 10^{-3}$ M$_\odot$ yr$^{-1}$ and $SFR<10^{-4}$ M$_\odot$ yr$^{-1}$ in TNG100-1 and TNG300-1, respectively \citep[see ][]{2019MNRAS.485.4817D}.

\begin{figure}
\centering
\includegraphics[trim = 6mm 6mm 5mm 5mm, clip, width=\columnwidth]{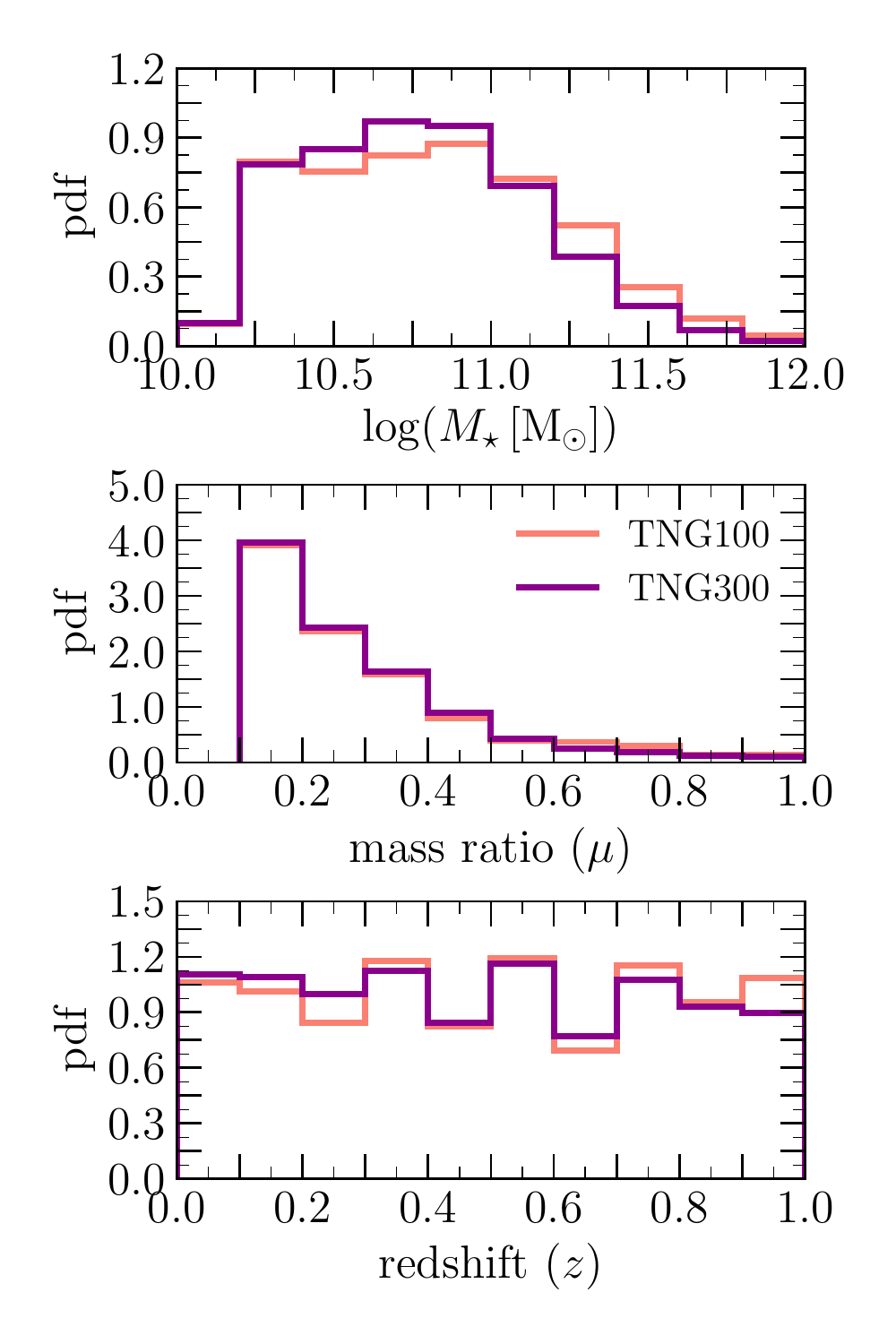}
\caption{The stellar mass ($M_\star$), mass ratio ($\mu$), and redshift ($z$) distributions of our post-merger samples selected from TNG100 (salmon histogram) and TNG300 (purple histogram). The sample spans a wide range in mass ratio ($\mu \ge 0.1$) and redshift ($z \le 1$) which is key when studying the effects of galaxy mergers during the post-merger stage. In total, we selected $1,855$ post-mergers from TNG100 and $25,836$ from TNG300.}
\label{fig:PM_summary}
\end{figure}

\begin{figure}
\centering
\includegraphics[width=\columnwidth]{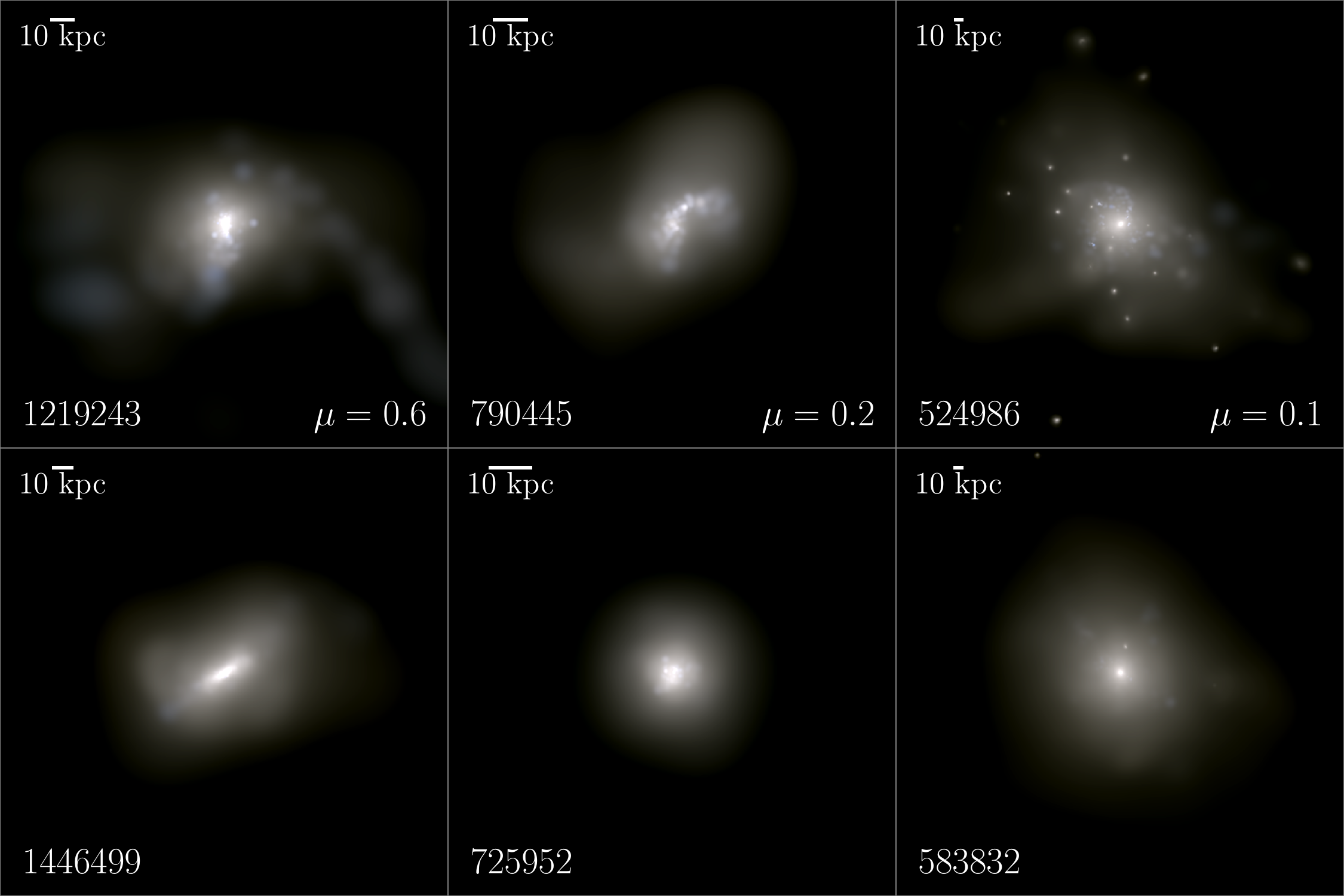}
\caption{An example of three $z=0$ post-mergers (top panels) with different mass ratios ($\mu$) along with their respective control galaxies (bottom panels) selected from TNG300-1. The panels show synthetic stellar composite images using the \texttt{JWST\_f200w, JWST\_f115w} and \texttt{JWST\_f070w} photometric filters. The \texttt{SubfindID} is indicated in the lower left corner of each panel, and scale bar (top left) indicates a $10$ kpc physical scale. The merger remnants exhibit evident low surface brightness features and disturbed morphologies. } 
\label{fig:PM_sample}
\end{figure}

Our post-merger sample consists of $1,855$ post-mergers in TNG100-1, and $25,836$ post-mergers in TNG300-1 with stellar masses $10^{10} \le M_\star / \mathrm{M}_\odot \le 10^{12}$, $z\le 1$, and $\mu \ge 0.1$. The properties of the post-merger sample are summarised in Figure \ref{fig:PM_summary} which shows the distributions of $M_\star$, $\mu$, and $z$ for the post-merger sample in TNG300-1 and TNG100-1. The post-mergers in our sample are uniformly distributed across redshift $0\le z \le 1$. While the merger sample in IllustrisTNG spans a large range of mass ratios, minor mergers ($\mu<0.3$) dominate the population of mergers. We note that the dearth of post-mergers at $M_\star \gtrsim 10^{10}$ M$_\odot$ is driven by the stellar mass limits applied in the post-merger selection (i.e., progenitor stellar mass $M_\star \ge 10^{10} $ M$_\odot$). Figure \ref{fig:PM_sample} depicts selected examples of post-mergers from our sample. The post-mergers exhibit disturbed morphologies with evident low surface brightness features (e.g., shells, tidal tails).

\begin{figure}
\centering
\includegraphics[trim = 6mm 6mm 6mm 5mm, clip, width=\columnwidth]{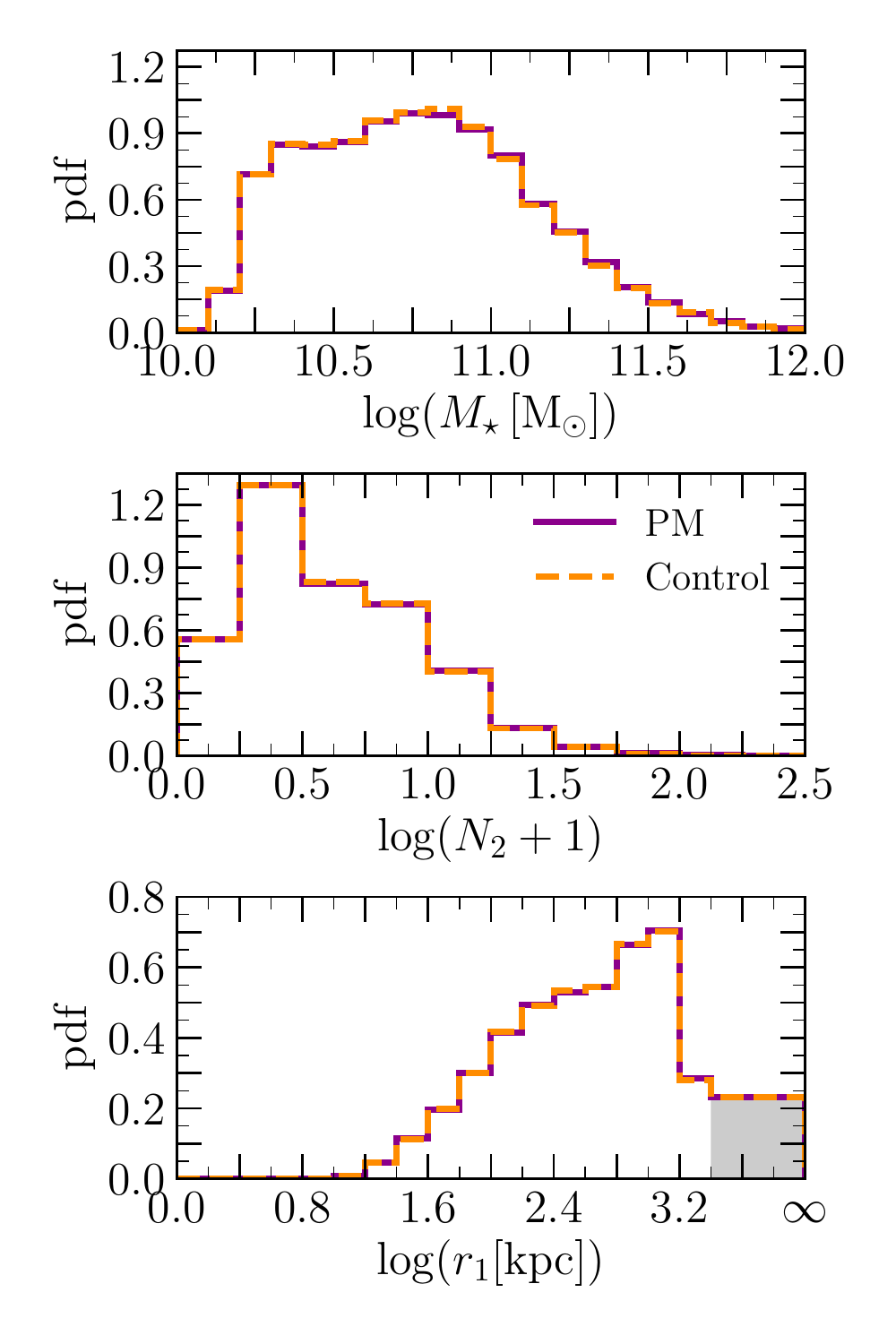}
\caption{A comparison between the post-merger sample and their respective controls in TNG300-1. The grey shaded bin in the $r_1$ distribution includes all galaxies with $r_1 > 2$ Mpc. The figure only shows the quantities which are allowed to vary during the control matching (i.e., $M_\star$, $N_2$, and $r_1$). The control matching is equally excellent for the TNG100 sample (not shown). }
\label{fig:control_summary}
\end{figure}

\subsection{Statistical controls}
\label{sec:methods:controls}
\noindent
We quantify the effects of galaxy mergers by adopting a commonly used control matching approach in observational studies of galaxy interactions/mergers \citep[e.g.,][]{2013MNRAS.435.3627E, 2013MNRAS.433L..59P}. For each post-merger, we identify a control galaxy to which we compare the properties of the post-merger. In this section, we describe the process of creating the control sample.

Large-box cosmological simulations provide a large and diverse population of galaxies thus allowing us to statistically study the changes in post-mergers compared to similar galaxies which have not undergone a merger. Comparing merger descendants to galaxies which have not undergone a recent merger allows us to isolate the effects of a merger whilst removing biases which may arise from known galaxy correlations (e.g., environment, stellar mass, redshift). 

Following \citet[]{2016MNRAS.461.2589P} and \citet{TNGpairs}, we assign a control galaxy to each post-merger in our sample. However, whereas \citet{TNGpairs} studied the pair interaction phase we are interested in studying post-mergers. Therefore, we apply some modifications to our control matching process. We first create a galaxy pool which includes all galaxies at $z\le 1$ with resolved SFRs, $M_\star \ge 10^{10}$ M$_\odot$, and $r_\mathrm{sep} > 2$. We exclude galaxies which have undergone a merger ($\mu \ge 0.1$) within the past $2$ Gyr thus generating an initial pool which includes $3,934,409$ galaxies in TNG300-1 ($263,794$ galaxies in TNG100-1). We then quantify the environment for each galaxy in our pool by calculating the number of neighbours within a $2$ Mpc radius (hereafter $N_2$) and the distance to the nearest neighbour with $M_\star \ge 0.1\times M_\mathrm{\star, host}$ (hereafter $r_1$). We also classify control galaxies and post-mergers as star-forming (SF) or passive. The classification is performed by applying a linear fit to the star-forming main sequence (SFMS) at stellar masses between $10^9 - 10^{10.2}$ M$_\odot$ which is extrapolated to higher $M_\star$ following the methodology of \citet{2019MNRAS.485.4817D}. Galaxies that lie below $2\sigma_\mathrm{MS}$ from the SFMS are classified as passive\footnote{$\sigma_\mathrm{MS}$ is the standard deviation of the SFMS residuals between $10^9 - 10^{10.2}$ M$_\odot$; for TNG300-1 $\sigma_\mathrm{MS}=0.3$ dex}. Finally, for each post-merger we select the single \textit{best} statistically matched control galaxy as follows: We first reduce the control pool to those galaxies that are in the same snapshot and have the same class (i.e., SF or passive) as the post-merger (i.e., matching in redshift and class). We stress the importance of the SF/passive class match: ignoring the class in the matching procedure would yield an unfair comparison where SF galaxies may have passive controls (and vice versa). Then, we search the snapshot-culled control pool for galaxies with $M_\star$ within a tolerance of $0.05$ dex, and $N_2$ and $r_1$ within 10\% of each post-merger. In most cases the initial tolerances yield at least one match. However, in some cases where no matches are found within the initial tolerances, we grow the tolerance range by $0.05$ dex and 10\% and repeat the search until at least one match is found. If more than one match is found, we select the \textit{best} match in all three parameters $M_\star$, $N_2$, and $r_1$ following the weighting scheme of \citet{2016MNRAS.461.2589P}. We exclude from our analysis post-mergers which require more than 3 grows to identify a control; four grows corresponds to unacceptably large matching tolerances (i.e., $\Delta \log M_\star = 0.2$, 40\% in $N_2$, and 40\% in $r_1$). Figure \ref{fig:control_summary} compares the distributions of post-mergers and their controls (both selected from TNG300-1) in the three matching parameters: $M_\star$, $r_1$, and $N_2$. The post-mergers are excellently matched by the control population. The TNG100-1 post-merger sample is matched to a similar quality as that of TNG300-1. 
 
Once the controls have been identified, we statistically quantify the enhancement (or suppression) in star formation by comparing post-mergers to their controls as an ensemble using the following metric:
\begin{equation}
    Q(sSFR) \equiv \frac{\langle sSFR_\mathrm{pm} \rangle}{\langle sSFR_\mathrm{control} \rangle}
    \label{eq:QsSFR}
\end{equation}
where $\langle sSFR_\mathrm{pm} \rangle$ is the arithmetic mean of the post-mergers' specific star formation rates (sSFRs), and $\langle sSFR_\mathrm{control} \rangle$ is the arithmetic mean of the controls' sSFRs. The sSFR is calculated using the SFR within twice the stellar half mass radius and the associated stellar mass. The metric used in this work has been extensively used in the literature to demonstrate the effects of galaxy interactions on star formation \citep[e.g., ][]{2013MNRAS.433L..59P, TNGpairs}.

%% file: results.tex
\section{Results}
\label{sec:results}
\noindent

\subsection{Effects of mergers}
\label{sec:results:effects}
\noindent

\begin{figure}
\centering
\includegraphics[trim = 5mm 6mm 5mm 5mm, clip, width=\columnwidth]{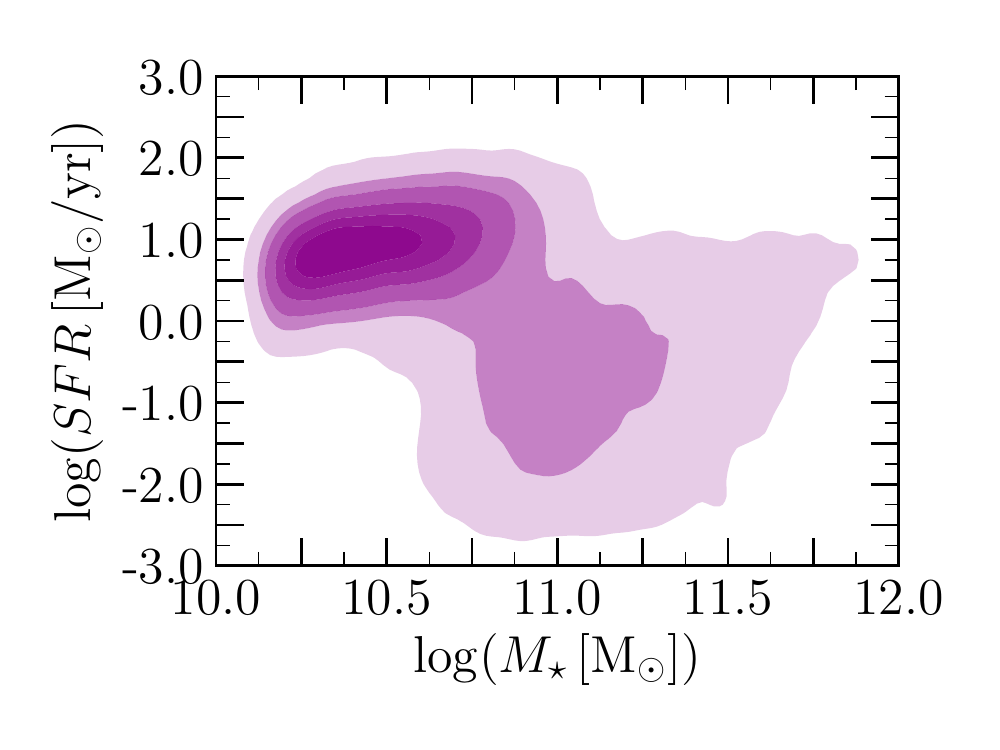}
\caption{The distribution of post-mergers in TNG300-1 in SFR$-M_\star$ space. The post-mergers sample we present spans a large range in SFR and stellar mass including star-forming galaxies as well as passive galaxies.}
\label{fig:PM_MS}
\end{figure}

%
\subsubsection{Star formation in post-mergers}
\label{sec:results:effects:delta_sSFR}
\noindent
We first investigate the effect of mergers on the SFR of the selected post-merger galaxy sample. Figure \ref{fig:PM_MS} shows the distribution of post-mergers in the SFR$-M_\star$ plane in TNG300-1, our fiducial simulation. While most of the post-mergers populate the star-forming galaxy main sequence ($67\%$ of the post-merger sample in TNG300-1), our post-merger sample spans a wide range in stellar mass and SFR including passive galaxies ($33\%$ of the post-merger sample in TNG300-1). Therefore, the post-merger sample selected from IllustrisTNG is well-suited for a systematic and statistical study of the SFRs in post-merger systems.

To \textit{fairly} investigate the effects of galaxy mergers on the post-mergers' SFR we compare the SFR of post-merger galaxies to their control counterparts. We remind the reader that the control galaxies are chosen to have not undergone a merger ($\mu\ge 0.1$) within the past $2$ Gyr while being matched to the post-merger sample in $N_2$, $r_1$, redshift, and $M_\star$. Table \ref{tab:QsSFR_summary} summarises the comparison of the average sSFRs of post-merger and their respective controls, $Q(sSFR)$ (defined in equation \ref{eq:QsSFR}). In the full sample, post-mergers statistically have elevated SFRs with a mean enhancement of $Q(sSFR) = 2.057 \pm 0.024$. 

The SFR enhancement (or lack thereof) is more pronounced for the star-forming and passive post-merger sub samples, respectively. The SFRs are enhanced in the star-forming sample with  $Q(sSFR) = 2.073 \pm 0.020$. The enhancement reported in our SF post-merger sample is consistent with those reported in observational studies. For example  \citet[]{2013MNRAS.435.3627E} measured a factor of 2.5 increase in the SFR of (star forming) post mergers in the SDSS, very similar to the factor of two found here. Conversely, the passive post-merger sample is statistically consistent with no SFR enhancement ($Q(sSFR) = 0.983 \pm 0.022$). 

In summary, the post-merger sample in IllustrisTNG exhibits a statistical enhancement in SFR  consistent with observations. The enhancements are more pronounced in star-forming post-mergers. In the following sections, we will dissect the dependence of the SFR enhancement/suppression on galaxy properties (e.g., redshift, stellar mass) and therefore gain insight on the physical mechanisms driving said enhancement/suppression.

\begin{table}
\begin{center}
\begin{tabularx}{1\columnwidth}{>{\raggedright\arraybackslash}X 
   >{\centering\arraybackslash}X >{\centering\arraybackslash}X }
\hline
\rule{0pt}{3ex}             & $Q(sSFR)$ & {\Large$\nicefrac{\sigma_\mathrm{Q(sSFR)}}{\sqrt{N}}$} \rule[-1.5ex]{0pt}{0pt}\\
\hline \hline
\rule{0pt}{3ex}all          & $2.057$   &    $0.024$ \\
\rule{0pt}{3ex}star-forming & $2.073$   &    $0.020$ \\
\rule{0pt}{3ex}passive      & $0.983$   &    $0.022$  
\rule[-1.5ex]{0pt}{0pt}\\
\hline
\end{tabularx}
\end{center}
\caption{A comparison of the sSFRs of post-mergers and their controls. The table reports $Q(sSFR)$ and the associated standard error on the mean for the full post-merger sample, and sub-samples therein. The post-merger sample generally has elevated $Q(sSFR)$ values. The signal is dominated by the star-forming post-mergers while the passive post-mergers exhibit sSFRs which are, on average, consistent with their controls.}
\label{tab:QsSFR_summary}
\end{table}

%

\subsubsection{Effect of redshift evolution}
\label{sec:results:effects:zred}
\noindent
From the perspective of galaxy mergers, the Universe was much sprier at earlier times: several theoretical and observational studies report the decline of the galaxy merger rate  with decreasing redshift \citep[e.g.,][]{2008ApJ...681..232L, 2009A&A...498..379D, 2011ApJ...742..103L, 2013A&A...553A..78L, 2015MNRAS.449...49R}. Whilst mergers can efficiently drive SFR enhancements at low redshifts, their contribution to the Universe's star formation rate density is thought to decrease with increasing redshifts \citep[e.g.,][]{2011ApJ...739L..40R, 2013MNRAS.429L..40K, 2014ARA&A..52..415M, 2017MNRAS.465.2895L, 2019ApJ...874...18W}. The scenario in which the contribution of galaxy mergers to star formation decreases with increasing redshift is challenged by other studies which report that galaxy interactions continue to induce star formation at high redshift \citep[e.g.,][]{2007ApJ...660L..51L, 2011ApJ...728..119W}. Therefore, to properly quantify the contribution of mergers to the global star formation rate density, one must assess the effect of mergers on SFR enhancements across cosmic time. For example, \citet{TNGpairs} examined the redshift dependence of SFR enhancement in galaxy pairs and reported smaller mean SFR enhancement with increasing redshift. \citet{TNGpairs} demonstrate that the decrease in the mean SFR enhancements at higher redshift is offset by the increased fraction of galaxies undergoing merger-triggered SFR enhancements leading to higher net SFR enhancements ($11\%$ at $z<0.2$ to $16\%$ at $0.6<z<1$) in their sample.

\begin{figure}
\centering
\includegraphics[trim = 5mm 6mm 6mm 6mm, clip, width=\columnwidth]{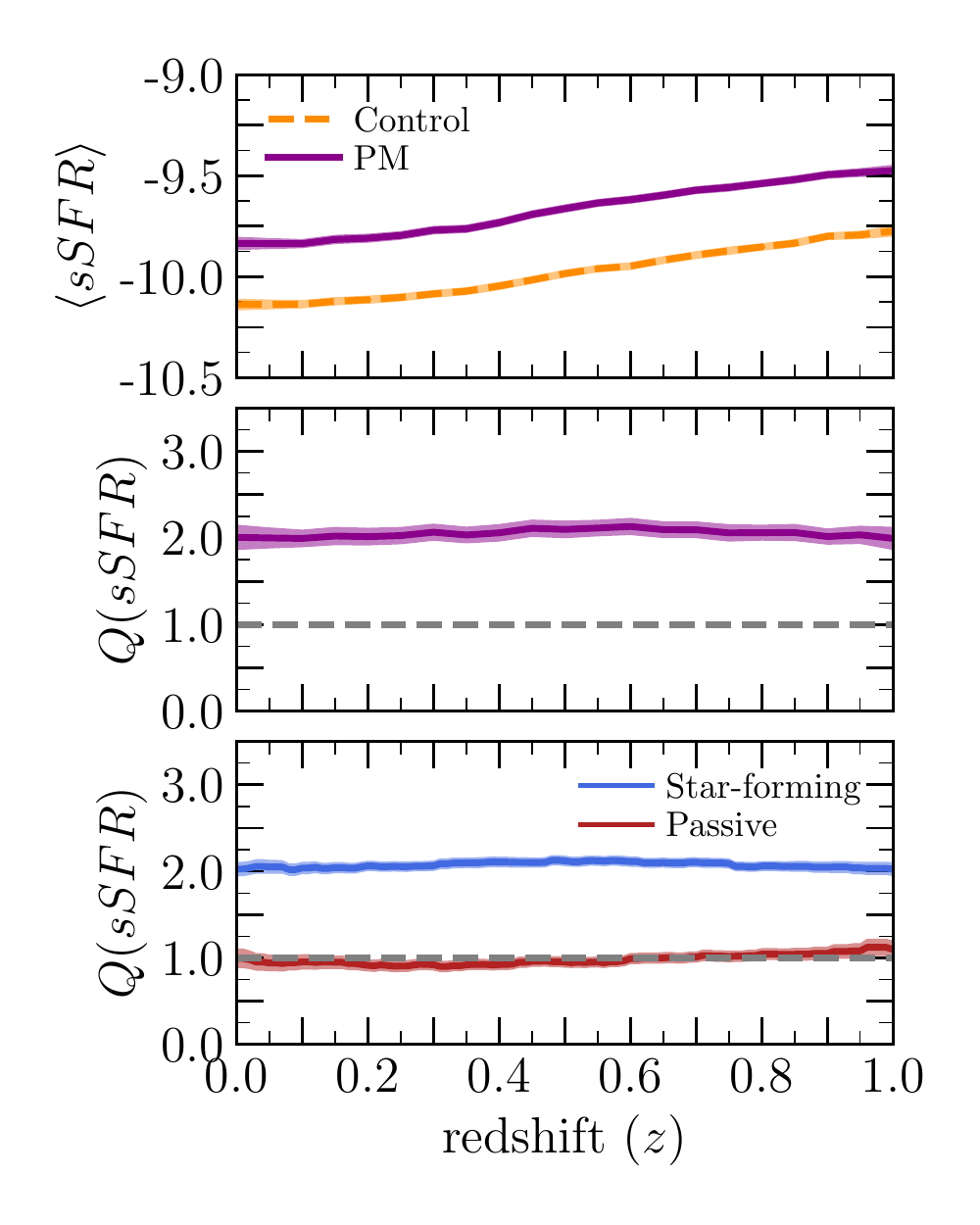}
\caption{The redshift-evolution of the sSFR enhancement in post-mergers. The top panel shows the running average sSFR for post-mergers (dark-purple) and their associated controls (orange) with the shaded region representing twice the standard error on the mean in bins or redshift. The middle panel shows the dependence of $Q(sSFR)$ (and the associated uncertainty; see equation \ref{eq:QsSFR}) on redshift for the full post-merger sample while the bottom panel shows $Q(sSFR)$ for star-forming (blue) and passive (red) post-mergers. Post-mergers exhibit enhanced sSFRs at all redshifts ($0\le z \le 1$). The enhancement is dominated by the star-forming post-mergers ($67\%$ of the sample) while the sSFRs of passive  post-mergers ($33\%$ of the sample) are consistent with those of the controls.}
\label{fig:Q_vs_zred}
\end{figure}

Figure \ref{fig:Q_vs_zred} shows the evolution in the mean sSFR across redshift for post-mergers and their controls (top panel), and the associated sSFR enhancement in post-mergers (middle and bottom panels). The mean sSFRs of both post-mergers and controls decline with decreasing redshift, a well-understood consequence of the decreasing accretion rate onto haloes \citep[e.g.,][]{2009ApJ...703..785D} and therefore the increasing quenched fraction \citep[e.g.,][]{2019MNRAS.485.4817D}. Despite the evolution of the average sSFR across cosmic time, post-mergers exhibit a steady sSFR enhancement, $Q(sSFR) \sim 2$, at all redshifts $0\le z \le 1$. By separating the sample by their SFRs, the lower panel of Figure \ref{fig:Q_vs_zred} shows that star forming post-mergers drive the aforementioned enhancement, while passive post-mergers have sSFRs which are statistically consistent with their respective controls. The distinction between the star-forming and passive post-merger samples hints at galaxy mergers enhancing the pre-existing conditions for star formation rather than triggering new processes; possibly emphasising the role of the galactic gas content (see \S \ref{sec:results:effects:fgas}).

Note that the sSFR enhancements driven by galaxy mergers persist consistently up to $z=1$ in our sample. This is consistent with results by \citet{2019MNRAS.490.2139R}, who analysed galaxy mergers with $\mu \ge 0.25$ in the \textsc{simba} simulation \citep{2019MNRAS.486.2827D} and found an enhancement of a factor of $\sim 2-3$ for $z\le2$. However, the absence of evolution in $Q(sSFR)$ with redshift in our sample is in contrast to recent work by \citet{TNGpairs}, who studied SFR enhancement during the pair phase using the same simulation suite and with similar methods as those used herein. \cite{TNGpairs} report a mild evolution in $Q(sSFR)$ with redshift: a mean $Q(sSFR)\sim 1.7$ at $0.8<z<1$ increases, with decreasing redshift, to $Q(sSFR) \sim 2.4$ at $z<0.2$. The different redshift evolution results between our study of post-mergers and \citeauthor{TNGpairs}'s pairs work may be due to differences in the galaxy properties of the two samples. Namely, the mass distribution of the post-merger sample presented here lacks post-mergers at $M_\star \sim 10^{10}$ M$_\odot$ while the pairs sample of \citet{TNGpairs} includes galaxies to lower masses. Alternatively, the difference in our control matching strategy and that of \citet{TNGpairs} may also contribute the aforementioned difference in redshift evolution. 

In order to more fairly compare our work to that of \citet{TNGpairs} we repeated our analysis using the exact matching strategy used in \citet{TNGpairs}. We find that the sSFR enhancement in the reconstructed post-merger sample evolves with redshift ($Q(sSFR) \sim 4$ at $z=0$ to $Q(sSFR) = 2.7$ at $z=1$). The redshift-evolution of the merger-induced SFR enhancement in the reconstructed star-forming post-mergers is even stronger, while the passive post-mergers now show suppressed SFRs (i.e., matched to star-forming control galaxies). Additionally, we tested the effects of the sample's mass distribution on the redshift evolution of $Q(sSFR)$. We repeated the analysis of \citet{TNGpairs} using our control matching strategy which resulted in no redshift evolution in $Q(sSFR)$. Therefore, the redshift dependence of $Q(sSFR)$ is driven by the control matching strategy: i.e., star-forming galaxies being matched to the passive galaxies which are more abundant at lower redshift.

%

\subsubsection{Effect of mass ratio}
\label{sec:results:effects:mu}
\noindent
Hydrodynamical simulations provide a solid understanding of the importance of the merger mass ratio on the measured enhancements and timescale on which the effects of a merger are visible \citep[e.g.,][]{2005A&A...437...69B, 2008MNRAS.384..386C, 2009ApJ...690..802J, 2010MNRAS.404..575L}. These previous works have found that minor mergers (small $\mu$) induce weaker starbursts, and short-lived observable asymmetries. On the contrary, major, gas-rich mergers induce the most prominent starbursts and the most long-lived asymmetrical features. However, previous works have mostly focused on mergers in idealised settings (i.e., binary disc mergers) where one can perform well targeted experiments to isolate the effects of galaxy properties (e.g., mass ratio, gas fraction) on the merger outcome (e.g., starburst strength, asymmetry metrics).

\begin{figure}
\centering
\includegraphics[trim = 5mm 6mm 6mm 6mm, clip, width=\columnwidth]{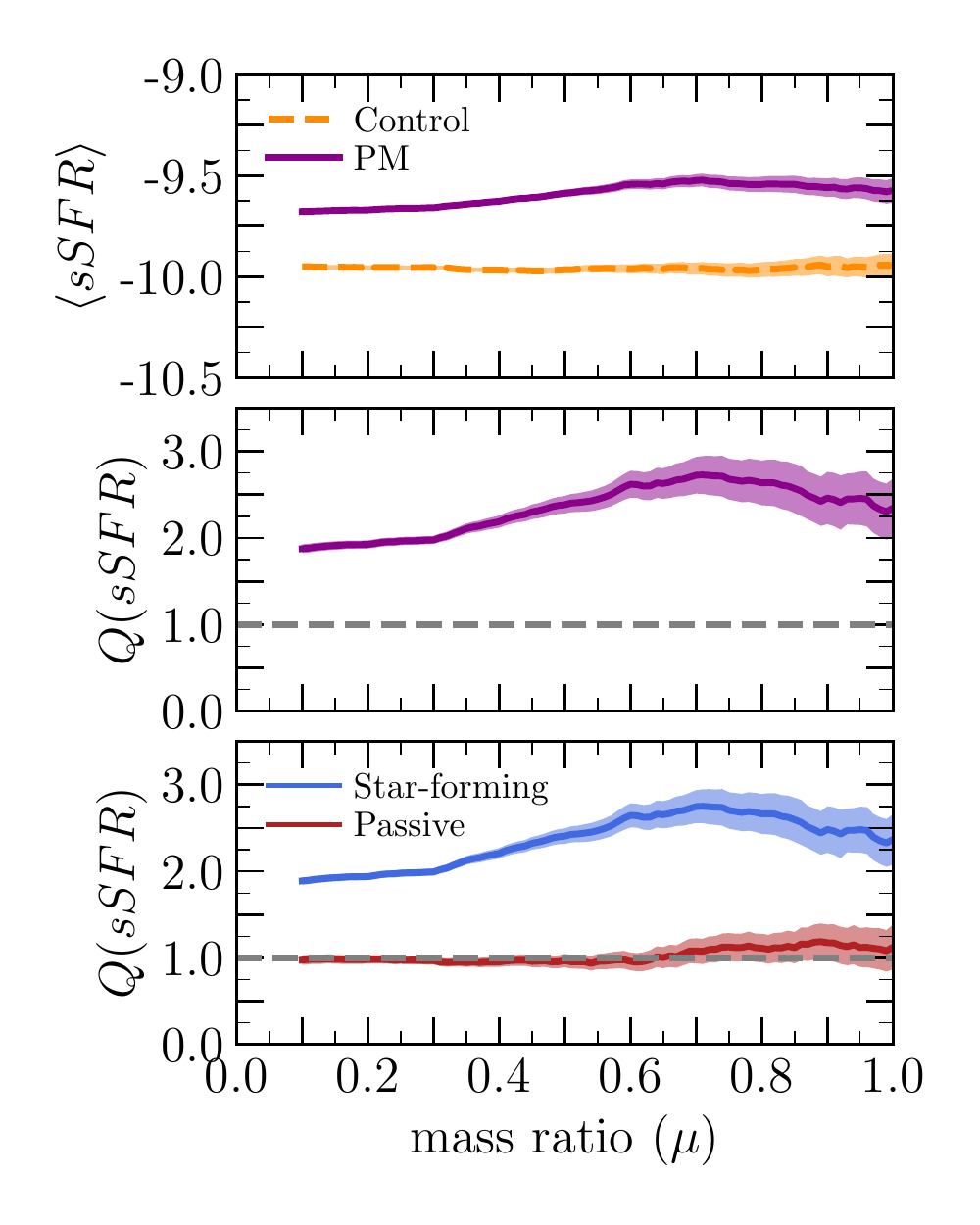}
\caption{The dependence of the enhancement in the post-mergers' sSFR on the parents' merger mass ratio ($\mu$). The shaded region represents twice the standard error on the mean in bins of $\mu$. Post-mergers exhibit enhanced sSFR for all mass ratios with an increasing enhancement for major mergers. The enhancement is dominated by the star-forming post-mergers, while the passive post-mergers sSFR are consistent with their controls.}
\label{fig:Q_vs_mu}
\end{figure}

In this sub-section, we investigate the relevance of the merger mass ratio ($\mu$) on the induced sSFR enhancement in a large-box cosmological simulation. Figure \ref{fig:Q_vs_mu} shows the mean sSFR at different mass ratios for post-mergers and their associated controls (top panel), and the dependence of the induced sSFR enhancement on the merger mass ratio (middle and bottom panels). Mergers with all mass ratios drive an enhancement in sSFR in our post-merger sample. The enhancement is more pronounced for major mergers ($\mu \ge 0.3$) with $Q(sSFR)\sim 2.5$, while minor mergers $\mu \sim 0.1$ show more modest (yet still significant) enhancements with $Q(sSFR)\sim 2$. Our results are consistent with other works exploring mergers in cosmological numerical simulations \citep[i.e.,][]{2019MNRAS.490.2139R}. Passive post-mergers exhibit sSFRs which are consistent with their controls while star-forming post-mergers dominate the enhancements shown in the full post-merger sample.

The results of the work presented here are qualitatively consistent with previous simulations of idealised mergers \citep[e.g.,][]{2008MNRAS.384..386C, 2009ApJ...690..802J}. Major mergers induce aggressive gravitational torques which drive dynamical instabilities in the ISM thus enhancing star formation. On the contrary, minor mergers induce more modest gravitational torques thus driving smaller enhancements. We note that the impact of minor mergers may be overestimated for galaxies which undergo multiple mergers; viz. if a galaxy undergoes a merger before the effects of a previous merger have decayed we would overestimate the boost in star formation. However, such a scenario is rare and the associated effects should be small.

\begin{figure}
\centering
\includegraphics[trim = 5mm 7mm 6mm 5mm, clip, width=\columnwidth]{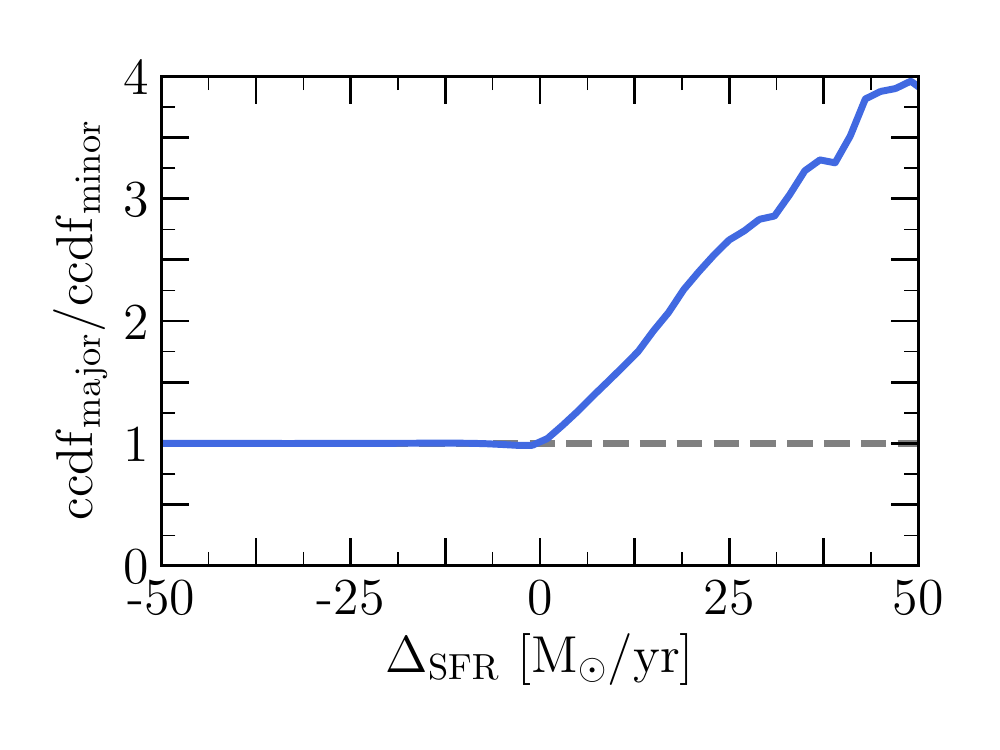}
\caption{The fraction of star-forming descendants of major mergers ($\mu \ge 0.3$) with SFR enhancement greater than $\Delta_\mathrm{SFR}$ (complementary cumulative distribution function; ccdf) relative to that of minor mergers ($0.1 \le \mu < 0.3$). The definition of $\Delta_\mathrm{SFR}$ is given in equation \ref{eq:delta_SFR}. Major mergers dominate the high tail of the $\Delta_\mathrm{SFR}$ distribution.}
\label{fig:BurstFrac}
\end{figure}

Comparisons to observational studies of post-merger galaxies are particularly complicated owing to the difficulty in assessing the mass ratio of a galaxy merger resulting in an observed post-merger galaxy. Nonetheless, one can qualitatively compare this work's results to the enhancements driven by galaxy mergers of varying mass ratios during the pair stage. For example, \citet{2012MNRAS.426..549S} studied galaxy pairs selected from SDSS and concluded that modest enhancements in SFR ($\log(SFR_\mathrm{pm}) - \log(SFR_\mathrm{control}) \le 0.45$ dex) can be achieved over a wide range of mass ratios ($0.1 \le \mu \le 1$), whereas the strongest SFR enhancements ($\log(SFR_\mathrm{pm}) - \log(SFR_\mathrm{control}) \ge 0.8$ dex) are preferentially driven by major mergers ($\mu \ge 0.3$). Figure \ref{fig:BurstFrac} shows the fraction of star-forming descendants of major mergers ($\mu \ge 0.3$) with enhancements greater than $\Delta_\mathrm{SFR}$  relative to that of minor mergers ($0.1 \le \mu < 0.3$). We define the merger-induced star formation, $\Delta_\mathrm{SFR}$, for a star-forming post-merger as the vertical offset in SFR from the SFMS: i.e., 
\begin{equation}
\begin{split}
\Delta_\mathrm{SFR} & = SFR_\mathrm{pm}- SFR_\mathrm{SFMS}.
\end{split}
\label{eq:delta_SFR}
\end{equation}
\noindent
Post-mergers below the SFMS contribute a negative $\Delta_\mathrm{SFR}$. Figure \ref{fig:BurstFrac} shows that strong enhancements of SFR are dominated by major mergers. For example, mergers that trigger an SFR that is $50$ M$_\odot$/yr above the SFMS are approximately four times as likely to be major, rather than minor, mergers.  Our results are therefore qualitatively consistent with the conclusions of \citet{2012MNRAS.426..549S}, namely that SFR enhancements are triggered by a wide range of mass ratios (Figure \ref{fig:Q_vs_mu}), but that the largest starbursts are preferentially produced by major mergers (Figure \ref{fig:BurstFrac}).

\begin{figure}
\centering
\includegraphics[trim = 5mm 6mm 6mm 5mm, clip, width=\columnwidth]{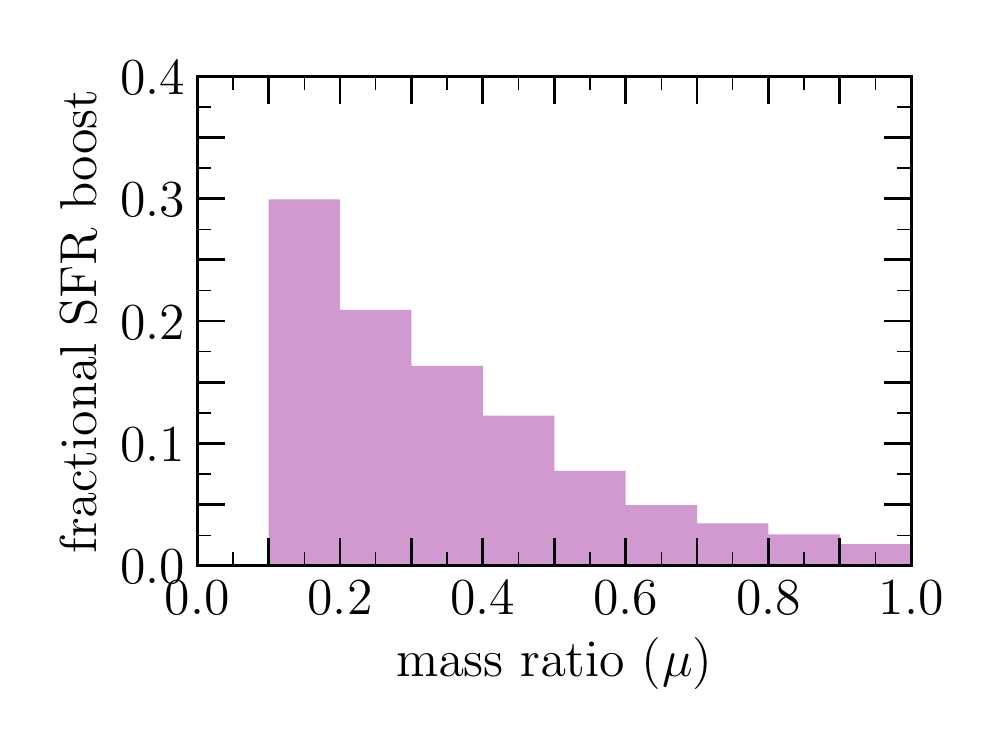}
\caption{An accounting of the contribution to the total merger-driven SFR enhancement budget in the star-forming post-merger sample. The merger-induced star formation boost for a given post-merger is defined as the vertical offset from the SFMS (see equation \ref{eq:delta_SFR}). The contribution to the total merger-induced SFR enhancement increases with decreasing mass ratio thus indicating that minor mergers ($0.1 \le \mu <0.3$) are significant contributors to the global merger-induced SFR budget. Although major mergers ($\mu \ge 0.3$) induce the strongest bursts, the abundance of minor mergers compensates for their modest SFR enhancement when compared to major mergers.}
\label{fig:boost_vs_mu}
\end{figure}

Although major mergers trigger the strongest starbursts, minor mergers are far more common (e.g. Figure \ref{fig:PM_summary}). In Figure \ref{fig:boost_vs_mu}, we therefore quantify the fractional contribution to the total merger-induced star formation budget\footnote{Note that we compute the total merger-induced SFR budget and \textit{not} the total cosmic star formation budget.} as a function of mass ratio in the star-forming post-merger sample. The contribution to the total merger-induced SFR enhancement budget increases with decreasing mass ratio. Despite their lower average enhancement, the cumulative enhancement from minor mergers ($0.1 \le \mu < 0.3$) accounts for $\sim 50 \%$ of the merger-induced SFR budget. Hence, minor mergers are a significant contributor to the merger-induced SFR budget. These results are qualitatively consistent with earlier idealised merger simulations \citep[e.g.,][]{2015MNRAS.449.3719S}. Since the contribution to the merger star formation budget continues to increase towards the smallest mass ratios in our samples (Figure \ref{fig:boost_vs_mu}), it would be of great interest to extend this study to even lower mass ratios. However, fully investigating the merger-induced SFR budget requires a much higher mass resolution. Such a study will be possible with the forthcoming IllustrisTNG50 simulation \citep{2019MNRAS.490.3234N, 2019MNRAS.490.3196P}.

\begin{figure}
\centering
\includegraphics[trim = 5mm 6mm 6mm 6mm, clip, width=\columnwidth]{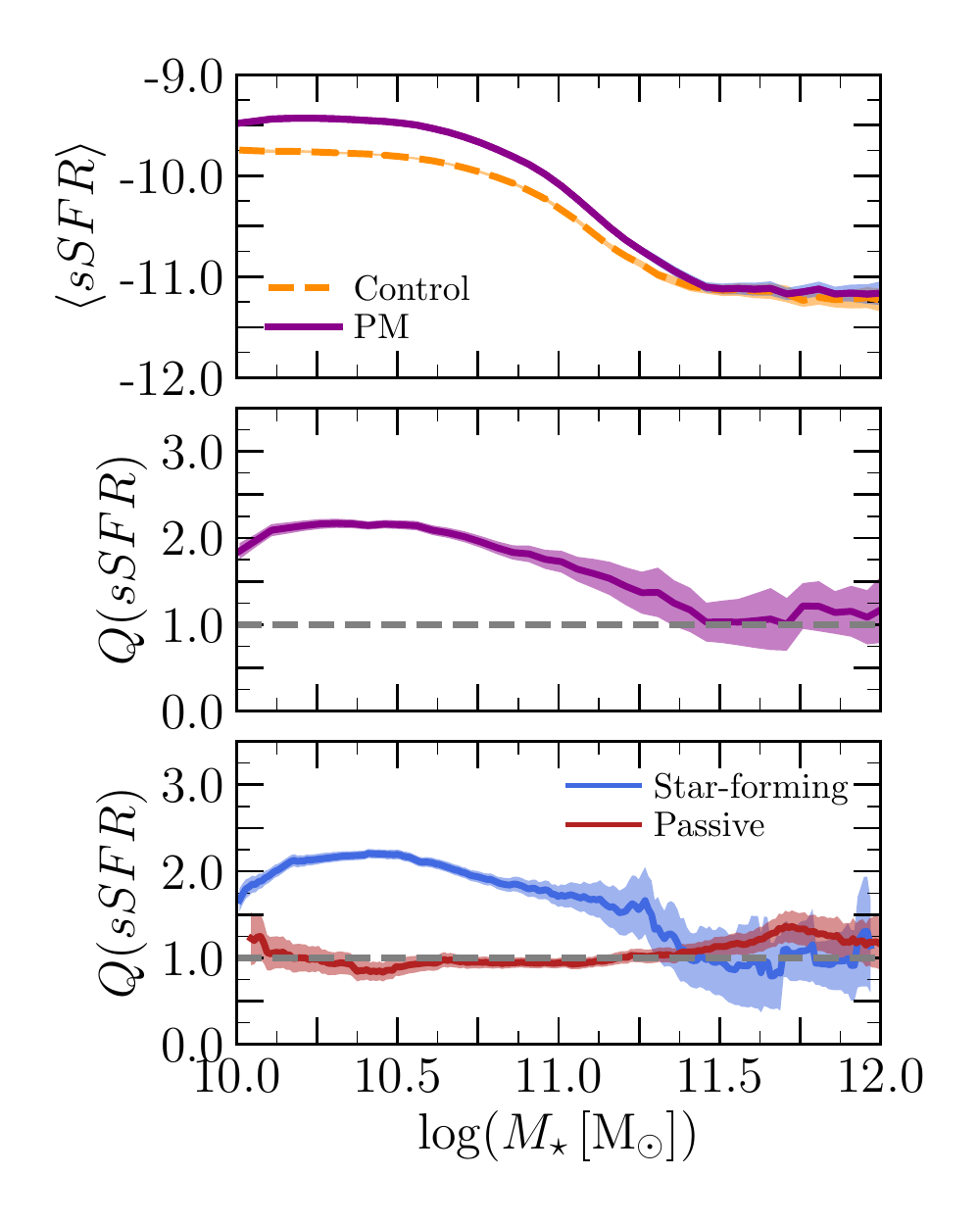}
\caption{The dependence of the enhancement in the post-merger's sSFR on post-merger stellar mass. The shaded regions represent twice the standard error on the mean in bins of $\log(M_\star/\mathrm{M}_\odot)$. Star formation is most enhanced in post-mergers with $10.0 \le \log(M_\star/\mathrm{M}_\odot) \le 11.4$ with a declining enhancement for larger $M_\star$. For $\log (M_\star / \mathrm{M}_\odot) > 11.4$ post-mergers form stars at an average rate which is consistent with that of the control galaxies. The star-forming post-mergers show the same behaviour as the full post-merger sample while passive post-mergers are consistent with no enhancement for all post-merger $M_\star$ with a possible slight enhancement at the largest $M_\star$.}
\label{fig:Q_vs_Mstar}
\end{figure}

%

\subsubsection{Effect of stellar mass}
\label{sec:results:effects:Mstar}
\noindent
In Section \ref{sec:results:effects:delta_sSFR} we demonstrated that the enhancement in sSFR is more pronounced in the star-forming population of the post-merger sample while the passive post-mergers show statistically consistent sSFRs with their controls. The discrepancy in sSFR enhancement between star-forming and passive post-mergers holds at all redshifts (Figure \ref{fig:Q_vs_zred}) and mass ratios (Figure \ref{fig:Q_vs_mu}). Knowing that there is a correlation between the fraction of passive galaxies and stellar mass (see Figure \ref{fig:PM_MS}), we investigate the dependence of the sSFR enhancement in post-mergers on stellar mass. We are interested in disentangling the source of the discrepancy in enhancement between star-forming and passive post-mergers.

Figure \ref{fig:Q_vs_Mstar} shows the mean sSFR of post-mergers and their controls (top panel), and the associated enhancement (middle and bottom panels) as a function of post-merger stellar mass. The average sSFR of post-mergers and controls declines with stellar mass reconfirming that passive galaxies dominate the high-mass galaxy population \citep[e.g.,][]{2010ApJ...719.1969B, 2014MNRAS.441..599B}. Unlike redshift and mass ratio, the stellar mass has a significant impact on the measured enhancement in sSFR. As the stellar mass increases, the sSFR enhancement diminishes to be consistent with the control sample and vanishes for $\log (M_\star /\mathrm{M}_\odot)>11.4$. Post-mergers with stellar masses $\log (M_\star /\mathrm{M}_\odot) \le 11.25$ exhibit an average enhancement in sSFR with $Q(sSFR) \sim 2$. On the contrary, the high stellar mass post-mergers, i.e. $\log (M_\star /\mathrm{M}_\odot) > 11.4$ are characterised by normal sSFRs (compared to their controls). The decline in $Q(sSFR)$ at $\log(M_\star /\mathrm{M}_\odot) < 10.25$ is caused by the minimum stellar mass limit we impose ($\log(M_\star /\mathrm{M}_\odot) \ge 10$) in order to reliably quantify the galaxies' recent merger histories: i.e., post-mergers with $\log(M_\star /\mathrm{M}_\odot) \sim 10$ will be paired with more massive controls which, on average, have higher SFRs thus leading to a decline in $Q(sSFR)$ at $\log(M_\star /\mathrm{M}_\odot) \sim 10$. The reported correlation of sSFR enhancement with stellar mass is consistent with the results of \citet{2019MNRAS.490.2139R} studying mergers in \textsc{simba}, another cosmological hydrodynamical simulation (with a different physical model).

Dissecting the post-merger sample further is key to understanding the source of the dependence of sSFR enhancement on stellar mass. The lower panel of Figure \ref{fig:Q_vs_Mstar} shows the enhancement at different stellar masses for the star-forming and passive post-merger sub-samples. While the star-forming post mergers follow the behaviour described above, the passive post-mergers exhibit typical sSFRs when compared to their controls at all stellar masses. Therefore, the stellar mass is not the fundamental driver of the differences between the star-forming and passive post-merger samples.\footnote{Similar conclusions are supported by Figures \ref{fig:Q_vs_zred}, and \ref{fig:Q_vs_mu}: The redshift and mass ratio are not fundamental drivers of the differences between the star-forming and passive post-merger samples.} This is suggestive of a separate galaxy property, which correlates with stellar mass and sSFR, driving the reported disparity between the passive and star-forming post-merger sub-samples (see \S \ref{sec:results:effects:fgas}).

\begin{figure}
\centering
\includegraphics[trim = 6mm 6mm 5mm 6mm, clip, width=\columnwidth]{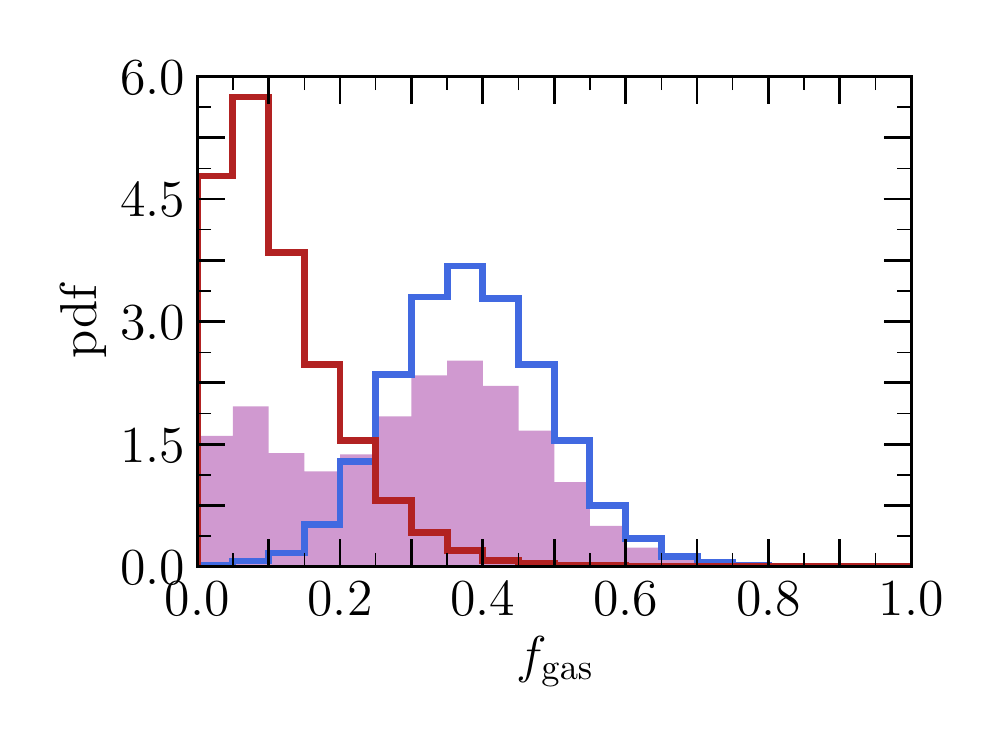}
\caption{The gas fraction distribution of the post-merger sample in TNG300-1. The gas fraction is defined as the progenitors' gas mass normalised by the total baryonic mass within $2\times R_\mathrm{half, \star}$ (see Equation \ref{eq:fgas}). The filled purple histogram represents the full sample while the red and blue histograms represent the passive and star-forming post-mergers, respectively. The star-forming post-mergers are the descendants of gas-rich mergers when compared to their passive counterparts. Note that $f_\mathrm{gas}$ accounts for all gas phases and therefore should not be compared to the molecular and cold gas fractions reported by observational studies \citep[e.g., ][]{2013A&A...550A..41C, 2018MNRAS.478.3447E, 2018ApJ...853..179T}.}
\label{fig:fgas_distro}
\end{figure}

\begin{figure}
\centering
\includegraphics[trim = 5mm 6mm 6mm 6mm, clip, width=\columnwidth]{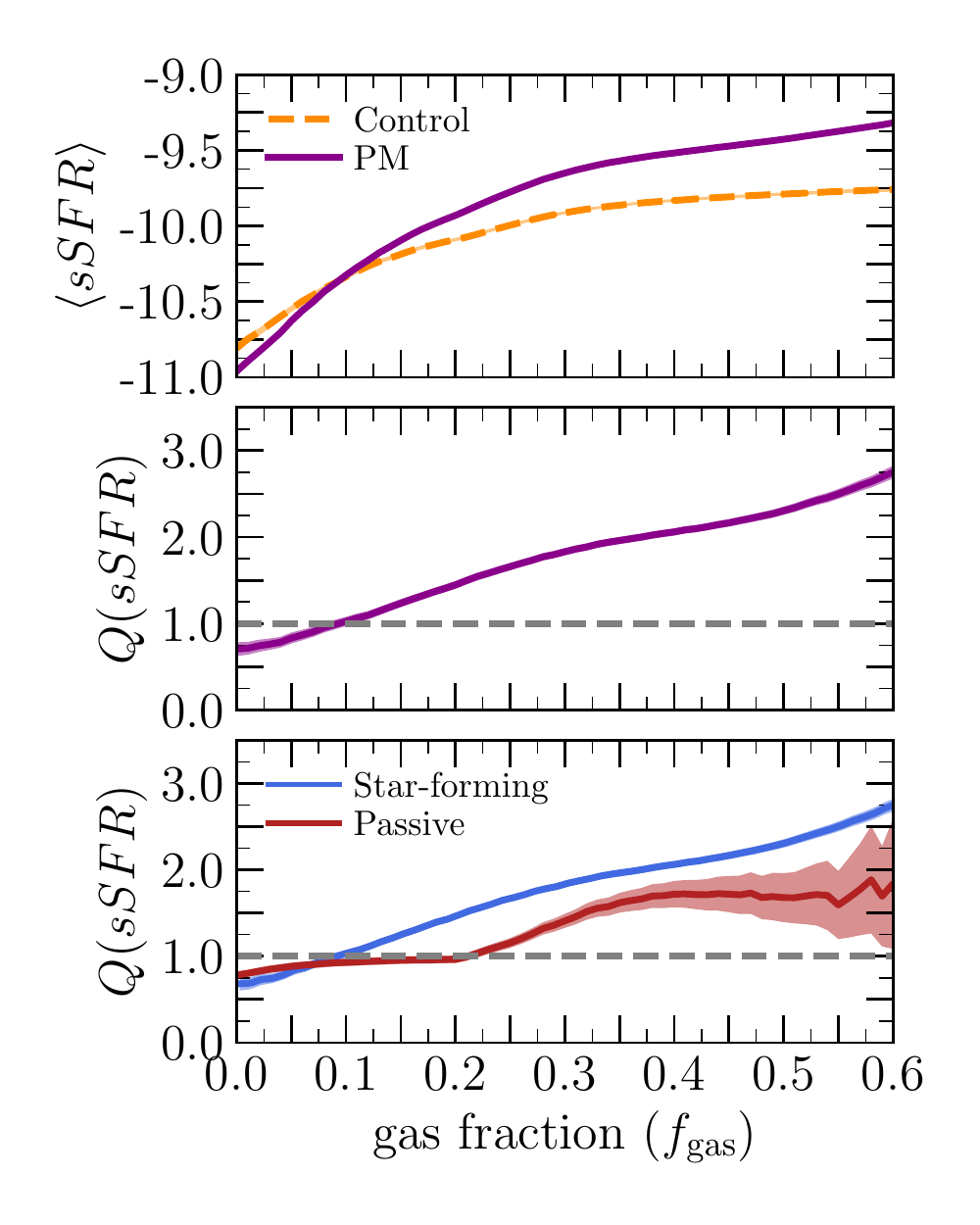}
\caption{The effect of the progenitors' gas content on the observed enhancement in the post-merger phase. The shaded regions represent twice the error on the mean in bins of $f_\mathrm{gas}$. Gas rich mergers (i.e., high $f_\mathrm{gas}$) yield post-mergers with enhanced sSFRs. Alternatively, mergers between gas-poor galaxies (i.e., low $f_\mathrm{gas}$) develop into post-mergers with suppressed sSFR. The aforementioned correlation between $Q(sSFR)$ and the parents' gas content holds for both star-forming and passive galaxies with the effects being especially pronounced for the star-forming post-merger sample. }
\label{fig:Q_vs_fgascont}
\end{figure}

%

\subsubsection{Effect of merger gas fraction}
\label{sec:results:effects:fgas}
\noindent 
The previous sections (\S \ref{sec:results:effects:zred}, \S \ref{sec:results:effects:mu}, and \S \ref{sec:results:effects:Mstar}) demonstrated that the redshift, mass ratio, and stellar mass may not be fundamental drivers of the discrepancy in the sSFR enhancement between passive and star-forming post-mergers. In fact, passive post-mergers show indistinguishable sSFRs from their controls at all redshifts, mass ratios, and stellar masses, while the star-forming post-merger exhibit strong enhancements at all $z$, $\mu$, and low $M_\star$. It is possible that the gas content, a fundamental contributor to star formation, is indeed driving the observed trends: gas content correlates with stellar mass, and drives star formation.

In this section, we explore the effect of the progenitor's gas content on the measured post-merger sSFR. We define 
\begin{equation}
f_\mathrm{gas} \equiv \frac{\displaystyle\sum_\mathrm{prog}M_\mathrm{gas}}{\displaystyle\sum_\mathrm{prog}M_\mathrm{gas} + \displaystyle\sum_\mathrm{prog}M_\mathrm{\star}}
\label{eq:fgas}
\end{equation}
\noindent
where $\sum_\mathrm{prog} M_\mathrm{gas}$ is the total gas mass (within $2\times R_\mathrm{half,\star}$) of the progenitors and $\sum_\mathrm{prog} M_\mathrm{star}$ is the total stellar mass (within $2\times R_\mathrm{half,\star}$) of the progenitors. Therefore, $f_\mathrm{gas}$ can be thought of as the total gas fraction available to form stars during the merger. Note that $f_\mathrm{gas}$ accounts for all gas phases and therefore should not be compared to the molecular and cold gas fractions reported by observational studies \citep[e.g., ][]{2013A&A...550A..41C, 2018MNRAS.478.3447E, 2018ApJ...853..179T}. Figure \ref{fig:fgas_distro} shows the distribution of $f_\mathrm{gas}$ for the post-mergers in our sample. The gas fraction distribution is bi-modal. While the post-merger sample includes descendants of both gas-rich and gas-poor mergers, star-forming post-mergers are predominantly the descendants of gas-rich mergers. 
The gas fractions in Figure \ref{fig:fgas_distro} are significantly higher than those reported in other studies using the IllustrisTNG simulations owing to differences in the definition of $f_\mathrm{gas}$. For example, \citet{2019MNRAS.490.3196P} report lower gas fractions for star-forming galaxies in the TNG50 box where they normalise the gas mass by the total dynamical mass within $2\times R_\mathrm{half, \star}$ instead of the total baryonic mass as shown in Equation \ref{eq:fgas}. We elect to normalise the gas mass by the baryonic mass following the commonly employed formalism in observational studies.

Figure \ref{fig:Q_vs_fgascont} shows the average sSFR for post-mergers and controls, and the associated enhancement as a function of $f_\mathrm{gas}$. As expected, both the sSFRs of post-mergers and controls correlate with $f_\mathrm{gas}$ (see the top panel of Figure \ref{fig:Q_vs_fgascont}). Additionally, the persistent enhancement reported in previous sections correlates strongly with $f_\mathrm{gas}$ with evident suppression in gas poor systems ($f_\mathrm{gas} < 0.1$), and elevated enhancements up to $Q(sSFR) \sim 3$ at the highest $f_\mathrm{gas}$. The correlation of $Q(sSFR)$ with $f_\mathrm{gas}$ also applies to the passive post-merger sub-sample, albeit to a lesser extent. Qualitatively, $Q(sSFR)$ for both passive and star-forming post-mergers show a consistent dependence on $f_\mathrm{gas}$: i.e., even passive post-mergers descending from gas rich mergers exhibit enhanced sSFRs when compared to their controls. 

We note that the level of enhancement (i.e., $Q(sSFR)$) of star-forming and passive post-mergers, and the critical value of $f_\mathrm{gas}$ above which enhancements are measured remain inconsistent between the star-forming and passive post-merger sample. Even at fixed $f_\mathrm{gas}$, the passive post-mergers have lower $Q(sSFR)$ (for most values of $f_\mathrm{gas}$), thus suggesting that the gas fraction is not the sole driver, albeit a strong driver, of SFR enhancement in galaxy mergers. Such a discrepancy could be engendered by various, possibly competing, effects such as: (1) the efficiency of feedback at different stellar masses may require higher $f_\mathrm{gas}$ for high $M_\star$ galaxies in order to achieve an enhancement; and (2) the metric $f_\mathrm{gas}$ does not differentiate between hot, cold, and molecular gas which may affect the quantitative results. In fact, the IllustrisTNG physical model treats star-forming gas using an effective equation of state. The sub-grid multi-phase pressurised ISM treatment does not follow the detailed properties and physical state of the gas (i.e., atomic and molecular fractions) which poses a limitation to further analysis of the gas content of post-mergers in this work. Particularly, with the current model, one cannot reliably discern the cause of the measured enhancement -- i.e., increased star formation efficiency or increased molecular gas content --  without further modelling of the state of the gas \citep[e.g.,][]{2011MNRAS.418.1649L, 2014MNRAS.440..920L, 2018ApJS..238...33D, 2019MNRAS.487.1529D} . Nonetheless, we will revisit the effect of the post-merger gas content in Section \ref{sec:discussion:matchMgas}.

In addition to the possible effects of feedback efficiency and the sub-resolution properties of the gas reservoir, galaxy morphology could contribute to the aforementioned discrepancy between star-forming and passive galaxies. Mergers of bulge-dominated galaxies have been predicted \citep[e.g.,][]{2008MNRAS.384..386C} and observed \citep[e.g.,][]{2012ApJ...758...73S} to have lower SFR enhancements. However, exploring the interplay between galaxy morphology and SFR enhancements in mergers within IllustrisTNG is beyond the scope of this work.

\begin{figure*}
\centering
\includegraphics[trim = 5mm 6mm 6mm 6mm, clip, width=\textwidth]{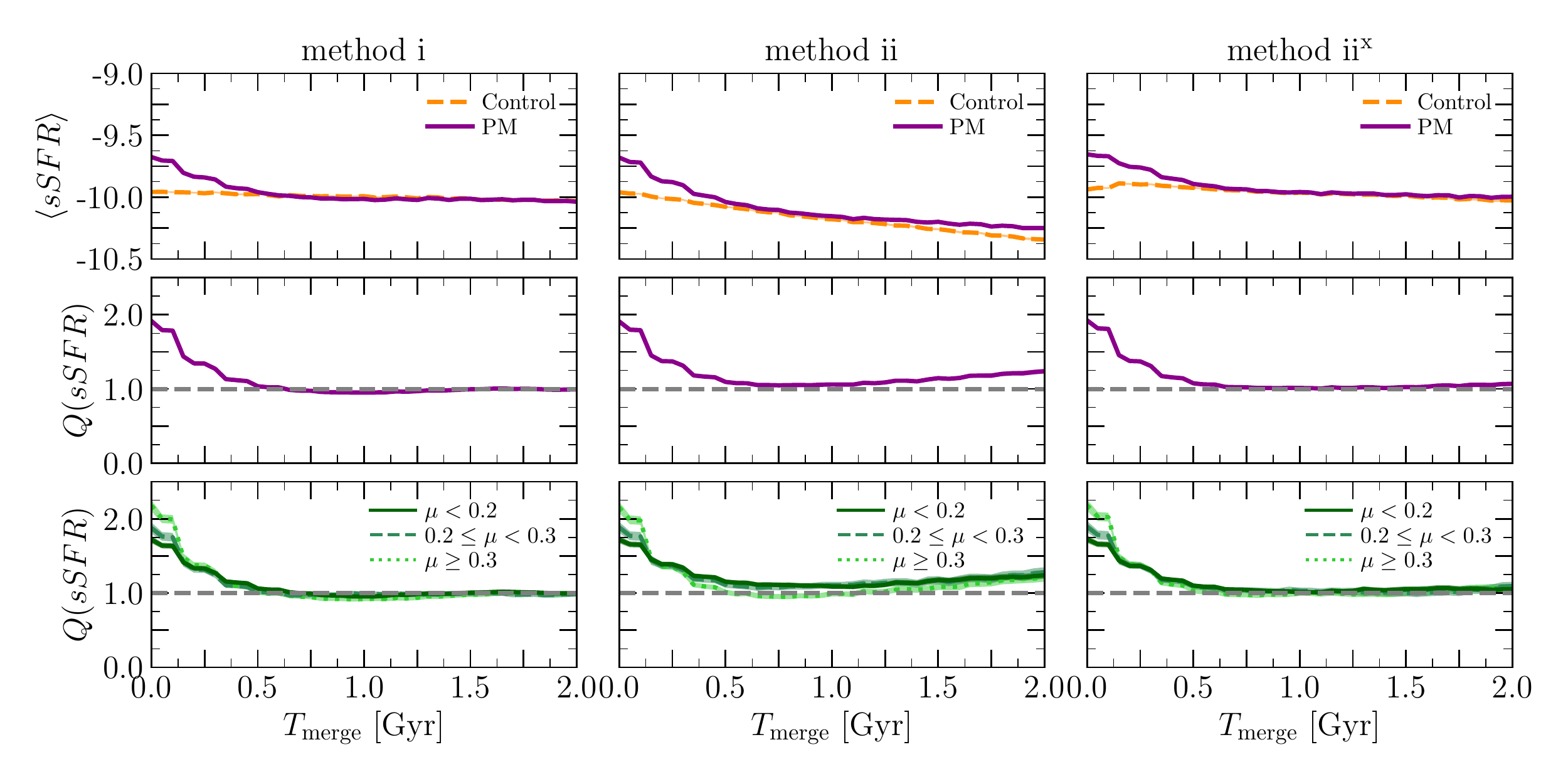}
\caption{The evolution of star formation during the post-merger phase. The three columns (from left to right) show the results of the tracing/control matching methods: method i, method ii, and method ii$^\mathrm{x}$, respectively. The top panels shows the running average sSFR for post-mergers (dark-purple) and their controls (orange) as a function of time after the merger. The middle panels show the evolution of $Q(sSFR)$ for the full post-merger sample, and the bottom panels show $Q(sSFR)$ for a sub-sample of post-mergers selected based on the parents' mass ratios. The shaded regions represent twice the standard error on the mean in bins of $T_\mathrm{merge}$. 
All methods show that the enhancement in sSFR decays following the merger and vanishes after $\sim 500$ Myr. Although mergers with different mass ratios induce enhancements of varying strengths, all enhancements decay similarly after $\sim 100$ Myr. 
The enhancement at later $T_\mathrm{merge}$ shown in method ii (middle column) is driven by the deviation in the post-merger and control properties (namely, different class for post-mergers and controls). Ensuring good control quality (method ii$^\mathrm{x}$; right column) removes the spurious enhancements seen at large $T_\mathrm{merge}$ in method ii. }
\label{fig:Q_vs_Tmerge__3methods}
\end{figure*}

\begin{figure}
\centering
\includegraphics[trim = 5mm 6mm 6mm 5mm, clip, width=\columnwidth]{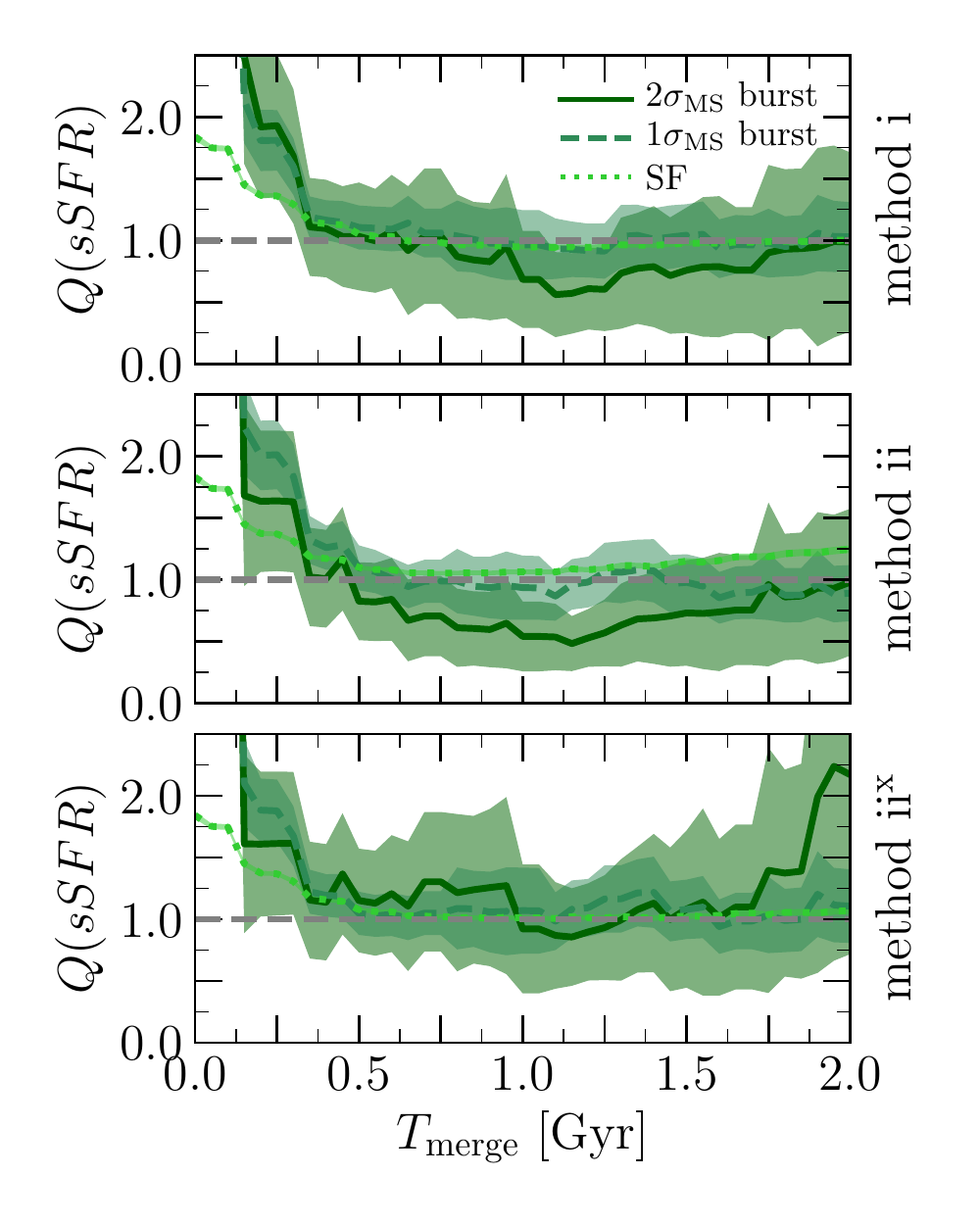}
\caption{The evolution of star formation during the post-merger phase for post-mergers with different merger-induced star-burst strengths. From top to bottom, the panels show $Q(sSFR)$ for a sub-sample of post-mergers selected based on the strength of the SFR at $T_\mathrm{merge}=0$ Gyr and traced using methods i, ii, and ii$^\mathrm{x}$, respectively. The shaded regions represent twice the standard error on the mean in bins of $T_\mathrm{merge}$. Independent of the method, the weakest starbursts (i.e., SF, $1\sigma_\mathrm{MS}$ burst) do not induce suppression in SFR. However, the strongest merger-induced SFRs can evolve to have statistically suppressed star formation. The level and existence of SFR suppression depend on the merger/control tracing method. Note: The $Q(sSFR)-$axis is truncated for visual purposes; $2\sigma$ bursts exhibit a peak $Q(sSFR)\simeq 12$  while $1\sigma$ bursts show an enhancement of $Q(sSFR)=8.4$ at $T_\mathrm{merge}=0$ Gyr. }
\label{fig:Q_vs_Tmerge__burst}
\end{figure}

\begin{figure*}
\centering
\includegraphics[trim = 7mm 9mm 7mm 7mm, clip, width=\textwidth]{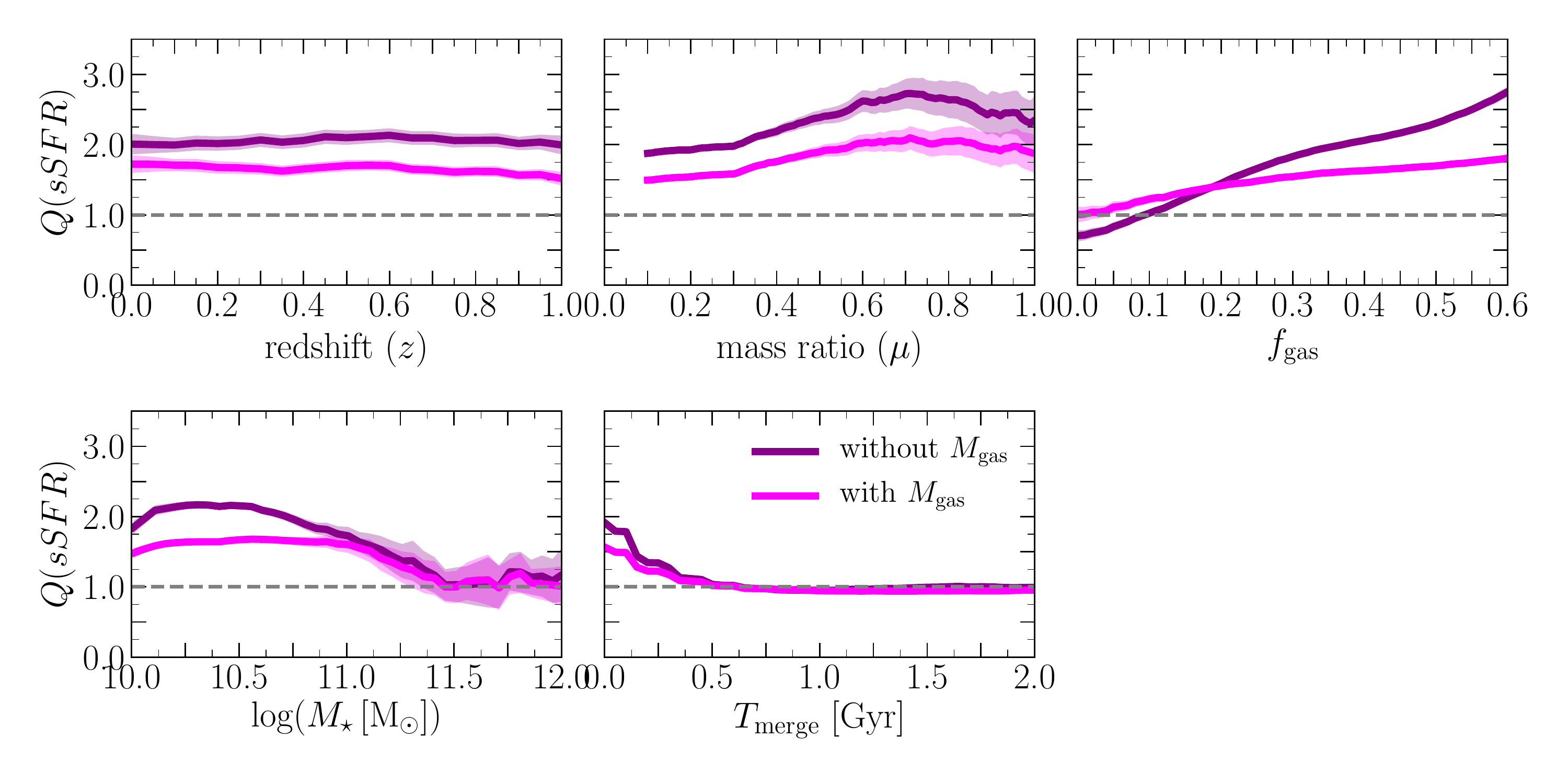}
\caption{The effect of the controlling for gas mass (i.e., gas fraction) on the measured merger-driven SFR enhancement. The shaded regions represent twice the standard error on the mean. The results of TNG300-1 for two different types of matching, with and without matching on gas mass, are shown in magenta and dark-purple, respectively. Controlling for the gas mass reduces the strength of the measured SFR enhancement (or suppression). Nonetheless the results are consistent between the two control matching techniques.}
\label{fig:Q_vs_X__Mgas}
\end{figure*}

\subsection{Evolution of post-mergers beyond coalescence}
\label{sec:results:evolution}
\noindent
In Section \ref{sec:results:effects} we demonstrated that post-mergers exhibit enhanced star formation rates compared to their controls, and tied the strength of the SFR enhancement to galaxy properties (e.g., $z$, $\mu$, $M_\star$, $f_\mathrm{gas}$). All the results presented thus far treat the post-mergers \textit{immediately} following the galaxy merger. In this section, we explore the forward evolution of the SFR enhancement (or suppression). 

We trace the evolution of the post-mergers in two distinct ways. 
\begin{enumerate}
    \item{\textbf{Our first method} tracks every post-merger forward in time until another merger occurs ($\mu \ge 0.1$) or we reach $z=0$. Then, at every snapshot, we identify control galaxies for the post-merger's descendants as described in Section \ref{sec:methods:controls}. Regenerating the controls independently for each descendant is akin to observational studies where we have no prior knowledge of a post-merger's parents or their properties.}
    \item{In contrast, \textbf{our second method} traces both the post-merger and the associated control (identified immediately after the merger) forward in time (similar to numerical simulations of isolated binary galaxy mergers) until either the control or the post-merger undergoes a merger with $\mu\ge 0.1$.}
\end{enumerate}
\noindent
We stress the subtle differences between the two methods. Calculating the controls at every snapshot independently compares the merger descendants to \textit{similar} galaxies. Therefore, our first method is insensitive to some of the effects of galaxy mergers (i.e., transitioning between star-forming and passive, quenching). For example, in our first method post-mergers and their controls belong to the same class, hence the effects of post-mergers transitioning between classes are suppressed. Additionally, post-mergers with unresolved SFRs are removed from our sample, thus our first method does not include the effects of quenching. On the other hand, comparing the merger descendants to the associated control's descendants at the same redshift highlights the difference in the evolution of post-mergers and the controls which evolve secularly. Therefore, our second method is sensitive to post-mergers changing class or quenching.

\subsubsection{Effects of the merger mass ratio}
\noindent
We first examine the evolution of the SFR enhancement and its dependence on the merger mass ratio. Figure \ref{fig:Q_vs_Tmerge__3methods} shows the evolution of sSFR and the sSFR enhancements over $2$ Gyr following the merger; $T_\mathrm{merge}$ is a measure of the time since the most recent merger with $\mu\ge 0.1$. The left column of Figure \ref{fig:Q_vs_Tmerge__3methods} depicts the results of independently regenerating the galaxy controls at each snapshot (method i). The merger-induced sSFR enhancement decays following the merger; after $\sim 0.5$ Gyr the post-mergers' sSFRs are consistent with those of the controls (left middle panel). The bottom left panel shows the evolution of $Q(sSFR)$ for different mass ratios computed using method i. While there \textit{is} a small dependence in the average $Q(sSFR)$ on $\mu$, the $\mu$-dependent enhancement only persists for $\lesssim 100$ Myr post-merger ($\sim 250$ Myr given the time resolution of the simulation). After $\sim 100$ Myr, the decay of $Q(sSFR)$ to normal values is identical for all mass ratios. 

The middle column of Figure \ref{fig:Q_vs_Tmerge__3methods} encapsulates the results of our second method. The controls at $T_\mathrm{merge} = 0$ Gyr are traced forward in time and compared to the post-merger descendants at their respective redshift. Similar to method i, the SFR enhancement in post-mergers decays over $\sim 0.5$ Gyr, and the dependence on the merger mass ratio is short lived. The upturn in $Q(sSFR)$ at large $T_\mathrm{merge}$ is caused by a deviation in the properties of the post-merger descendants and control's descendants which renders the control's descendants inadequate control matches for the post-merger descendants. Namely, the post-merger descendants and the control descendants do not belong to the same class (i.e., star-forming, passive) which leads to an unfair comparison of galaxies and their SFRs. 

In order to mitigate the effects of the control quality on the results, we introduce a refined version of method ii, hereafter method ii$^\mathrm{x}$. We remove post-merger descendants (and their associated controls) if the control quality does not adhere to the conditions described in section \ref{sec:methods:controls}. Namely, we remove post-merger descendants (and their associated controls) in cases of unacceptably large matching tolerances (i.e., $\Delta \log (M_\star) > =0.2$, 40\% in $N_2$ and $r_1$), unresolved SFRs, and when the post-mergers and their associated controls do not belong to the same class (i.e., star-forming, passive). The results are shown in the right column of Figure \ref{fig:Q_vs_Tmerge__3methods}. Constraining the quality of control matching (particularly the control and post-merger class) alleviates the spurious enhancement at $T_\mathrm{merge} > 1$ Gyr while maintaining similar results to method i (decay timescale of the SFR enhancement, and the short-lived dependence on $\mu$). 

All our methods show that the SFR enhancement decays following the merger and vanishes after $\sim 500$ Myr, with little dependence on mass ratio. Although the mass ratio has been reported to have a significant impact on the extent and time-scale of observed morphological disturbances of post-mergers \citep[i.e.,][]{2010MNRAS.404..575L}, our post-merger sample indicates that the evolution of SFR enhancement beyond $\sim 100$ Myr may be independent of mass ratio. Nonetheless, the decay timescales presented in this work are broadly consistent\footnote{Comparing the results presented in this work with idealised binary galaxy merger simulations is not wholly fair. Idealised simulations traditionally use gas-rich disc galaxies which does not reflect the nature of the post-mergers in IllustrisTNG.} with results from high resolution idealised simulations \citep[e.g.,][]{2015MNRAS.448.1107M}. We stress a possible caveat which may affect our results: Since $T_\mathrm{merge}$ measures the time since the most recent merger, it does not have any memory of a galaxy's previous history. Therefore, consecutive mergers (although rare) may hinder our ability to discern the effects of mergers with different mass ratios.

\subsubsection{Effects of the merger-induced starburst strength}
\noindent 
We next investigate the effect of merger-induced starbursts on the evolution of post-merger galaxies. Figure \ref{fig:Q_vs_Tmerge__burst} shows the evolution of $Q(sSFR)$ for post-mergers with different SFR strengths at $T_\mathrm{merge} = 0 $ Gyr: starburst galaxies with $SFR > SFR_\mathrm{MS} + 2\sigma_\mathrm{MS}$, starburst galaxies with $SFR > SFR_\mathrm{MS} + 1\sigma_\mathrm{MS}$, and the rest of the star-forming post-mergers. The top, middle, and bottom panels show the results of method i, method ii, and method ii$^\mathrm{x}$, respectively.
Independent of the method, the weakest starbursts (i.e., star-forming galaxies, or galaxies with $SFR > SFR_\mathrm{MS} + 1\sigma_\mathrm{MS}$) never lead to a statistical reduction in SFR following the merger. However, some mergers (with the strongest merger-induced SFRs) can evolve to have statistically suppressed star formation, although this is dependent on the control matching method. 
For example, method i (new controls generated at each snapshot) is insensitive to galaxy class transitions or quenching. Therefore, no SFR suppression is evident shortly after the merger. However, at $\sim 1$ Gyr, the strongest bursts evolve to have slightly suppressed SFRs. The suppression is caused by the systematically lower SFRs (compared to typical star-forming controls) of previously passive post-merger descendants evolving to be star-forming, an effect previously reported as `rejuvenation' by \citet{2019MNRAS.490.2139R}. The suppression vanishes after $\sim 2$ Gyr. 
In method ii (the same controls are traced in time alongside the post-mergers), the strongest merger-induced starbursts evolve to have statistically suppressed SFRs compared to their secularly evolving controls beyond $\sim 0.5$ Gyr. After $\sim 2$ Gyr, the merger descendants have SFRs that are consistent with their controls. We note that method ii provides an unfair comparison of star-forming galaxies to passive or quenched (unresolved SFRs) galaxies which causes the spurious enhancements at late $T_\mathrm{merge}$ (see Figure \ref{fig:Q_vs_Tmerge__3methods}). 
Method ii$^\mathrm{x}$ alleviates the unfair comparison in method ii. Consequently, forcing the post-merger descendants and the control descendants to belong to the same class conceals the suppression evident in method ii. The dependence of the SFR suppression on class matching suggests that merger descendants exhibiting merger-induced starbursts evolve not only to have statistically suppressed SFRs but also become passive or quenched on faster timescales than their controls.

Our sample shows that the descendants of galaxy mergers can evolve to have modestly suppressed SFRs that may persist for a Gyr or so. Some post-merger descendants can be rejuvenated after a period of SFR suppression. However, the level and existence of SFR suppression depends on the control matching method (unlike the SFR enhancements which are very robust to method). The results presented here pertaining to merger-induced starbursts are consistent with the framework where galaxy mergers may drive SFR suppression \citep[e.g.,][]{2008ApJS..175..356H}. However, a detailed exploration of the connection between galaxy mergers and SFR suppression (i.e., timescales, physical processes, feedback) is beyond the scope of this work. By taking the different methods we have 1) shown that the largest starbursts can eventually suppress the SFR where the descendants become passive or quenched faster than their controls, but 2) the techniques used in observations (equivalent to method i and method ii$^\mathrm{x}$) would be largely insensitive to suppression in SFR.

%% file: discussion.tex
\section{Discussion}
\label{sec:discussion}
\noindent

\begin{figure*}
\centering
\includegraphics[trim = 7mm 9mm 7mm 7mm, clip, width=\textwidth]{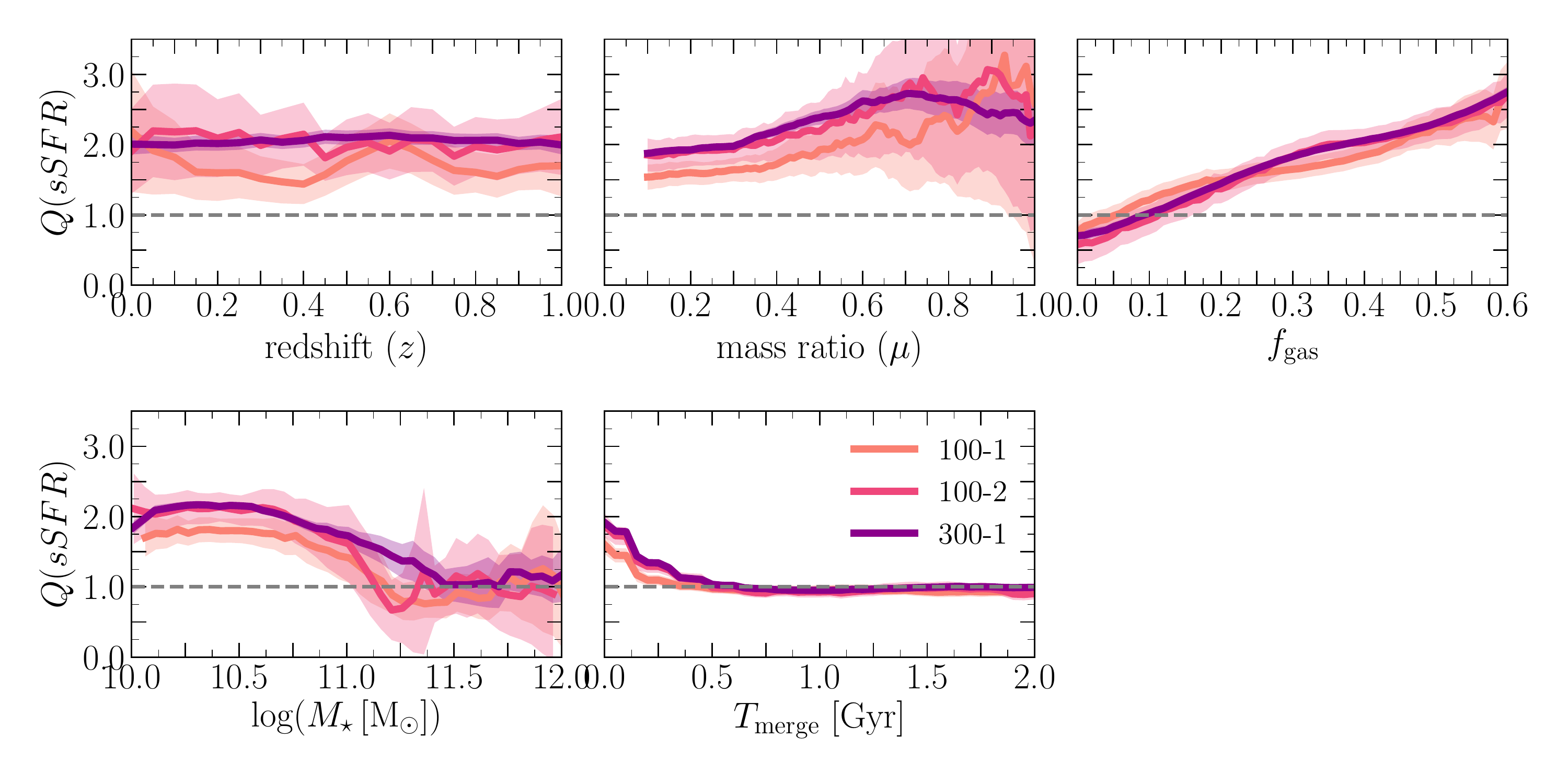}
\caption{The effect of the simulation resolution on the results presented in this work. The shaded regions represent twice the standard error on the mean. The results of TNG100-1, TNG100-2, and TNG300-1 are broadly consistent. The effect of resolution are most prominent in the timescale on which the $Q(sSFR)$ decays in the post-merger phase. The enhancement in the lower resolution simulations (i.e., TNG100-2 and TNG300-1) wanes over $\sim 500$ Myr vs. $\sim 300$ Myr for the higher resolution simulation (i.e., TNG100-1).}
\label{fig:Q_vs_X__Res}
\end{figure*}

\subsection{Controlling for gas fraction}
\label{sec:discussion:matchMgas}
\noindent
In Section \ref{sec:results:effects:fgas} we showed that the SFR enhancement correlates strongly with the progenitors' gas fraction ($f_\mathrm{gas}$). Yet, our results do not account for the post-merger's gas content when searching for a control galaxy. Understanding the effect of controlling for the gas content on the presented results is a vital test for observational studies, because the choice of matching parameters can significantly affect the conclusions. For example, \citet{2015MNRAS.449.3719S} report different correlations between SFR and gas fraction depending whether they control for gas fraction; i.e., the merger-induced SFR enhancement correlates with the gas fraction \textit{only} when the sample is \textit{not} matched in gas fraction. In Figure \ref{fig:Q_vs_fgascont} we already showed that, in agreement with \citet{2015MNRAS.449.3719S}, there is a correlation between SFR enhancement and gas fraction when gas fraction is not included in the control parameters. Now, we investigate whether the aforementioned result changes if gas fraction is controlled for. 

The control sample is regenerated using the single best match in $M_\mathrm{gas}$, $M_\star$, $z$, $N_2$, and $r_1$ as described in Section \ref{sec:methods:controls}. An initial tolerance of $0.05$ dex is used for $M_\mathrm{gas}$; if no matches were found, we grow the tolerance by $0.05$ dex. All the other parameters are treated as before (see \S \ref{sec:methods:controls}).

Figure \ref{fig:Q_vs_X__Mgas} shows a summary of the merger-driven SFR enhancements, initially reported in Section \ref{sec:results}, re-calculated using the $M_\mathrm{gas}$-matched control sample. The results remain qualitatively unchanged although the strength of the enhancement (or suppression) is reduced when controlling for $M_\mathrm{gas}$. The results presented here are consistent with the results of \citet{2016ApJS..222...16C}, who report enhanced SFRs in their star-forming spiral pairs compared to control galaxies with similar gas mass.

In terms of a direct comparison with observations, it is worth noting that the IllustrisTNG physical model does not explicitly track the atomic and molecular fractions in gas cells during the simulation. Therefore, we are unable to discern the sub-resolution physical state (i.e., atomic fraction, molecular fraction) of the gas reservoirs in the post-mergers and their controls. However, knowing that the IllustrisTNG physical model attributes star formation to gas cells which meet the sub-resolution star formation criterion ($SFR \propto \rho^{1.5}$), one can use the SFR as a proxy for the molecular gas content of galaxies \citep[see][]{2019MNRAS.487.1529D}. The measured SFR enhancement, which persists even when controlling for gas fraction, is suggestive of an enhanced \textit{molecular gas} content in post-mergers. We will explore gas fraction evolution during galaxy mergers more extensively in the future. Our results are qualitatively consistent with observational studies reporting that post-mergers are home to an enhanced molecular gas content \citep[e.g.,][]{2018ApJ...868..132P, 2018MNRAS.476.2591V}. Additionally, our results are broadly consistent with the high resolution simulations of \citet{2019MNRAS.485.1320M} who showed that the cold-dense gas content of galaxies is enhanced during the pair phase and immediately before the coalescence.

\begin{figure*}
\centering
\includegraphics[trim = 5mm 6mm 6mm 6mm, clip, width=\textwidth]{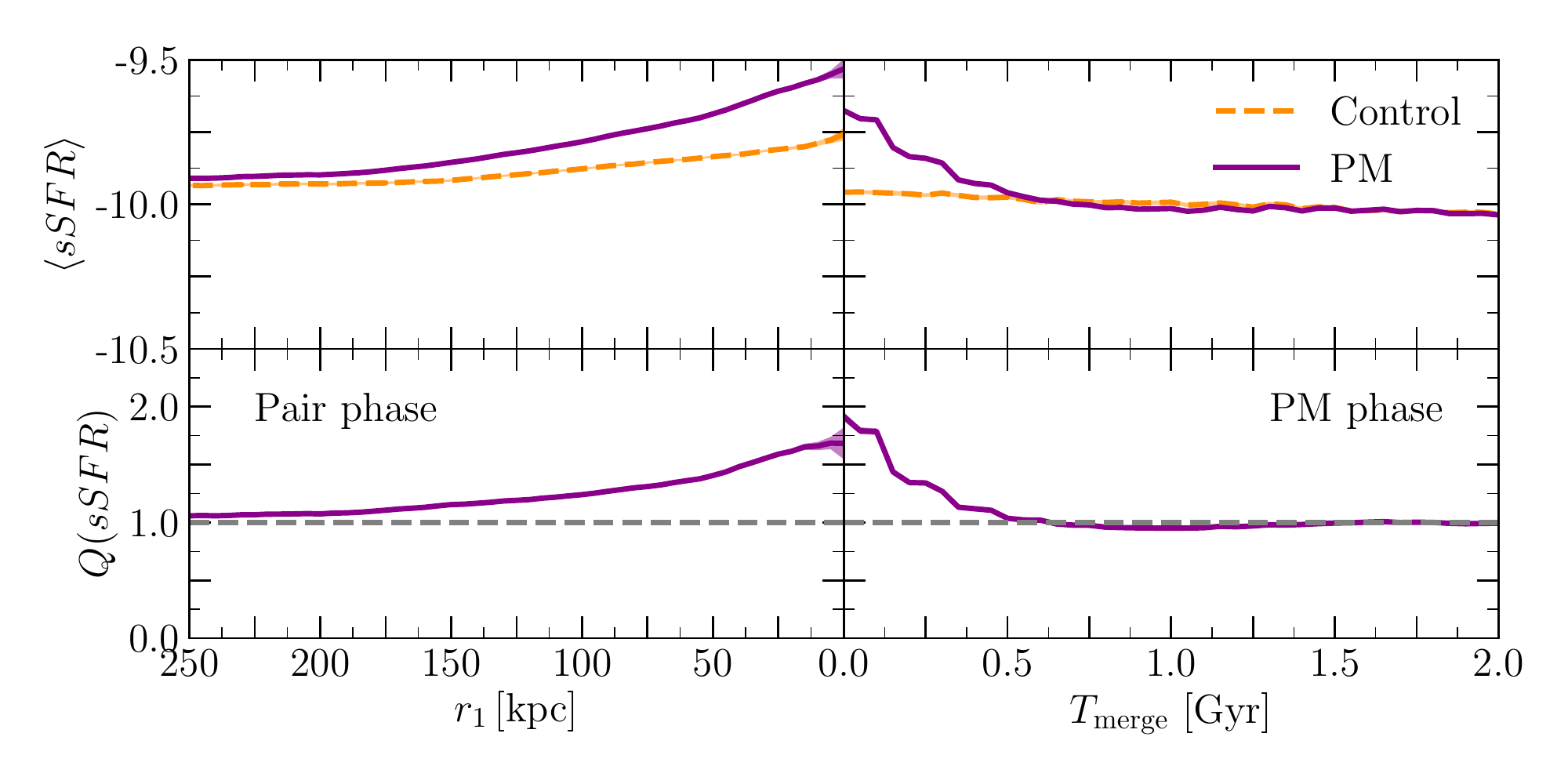}
\caption{Reconciling the observable evolution of mergers during the galaxy pair phase \citep[left panels; ][]{TNGpairs} and the evolution of post-merger systems (right panels; see Figure \ref{fig:Q_vs_Tmerge__3methods}) in TNG300-1. The controls are recalculated independently for each merger remnant as in the left column of Figure \ref{fig:Q_vs_Tmerge__3methods} (i.e. method i). The shaded regions (although remarkably small) represent twice the standard error on the mean. As the separation of galaxy pairs ($r_1$) decreases, the galaxies experience an enhancement in their SFRs. Following coalescence, the SFR enhancement decays to be consistent with the controls.}
\label{fig:Q_vs_r1-et-Tmerge}
\end{figure*}

\subsection{Effects of simulation resolution}
\label{sec:discussion:restest}
\noindent
A common effect of numerical resolution on star formation arises from the correlation of star formation with dense gas: i.e., stars are spawned from gas cells above a given density threshold (see \S \ref{sec:methods}). As the simulation's resolution (mass and spatial) increases, we are able to sample progressively denser gas. Therefore, more gas cells are eligible to form stars which increases the star formation rate. However, the effects of the increased star formation (and associated feedback) are overly complex. In this section, we perform a convergence test to determine how robust our results are to changes in the simulation's resolution. We direct the reader to \cite{2018MNRAS.473.4077P} for a general discussion on the convergence of the IllustrisTNG physical model. 

The results reported in Section \ref{sec:results} are for TNG300-1. Although TNG300-1 has a lower spatial resolution compared to TNG100-1, TNG300-1 provides a large number of post-mergers which allows for the statistical analysis of the post-merger suite that is necessary when exploring SFR enhancement as a function of other parameters. We use three simulations to test the effects of numerical resolution: TNG100-1, TNG100-2, and TNG300-1. See Section \ref{sec:methods} for a description of each of the three simulations. Figure \ref{fig:Q_vs_X__Res} summarises the results of Section \ref{sec:results} for all three simulations. The simulations with similar resolution (i.e., TNG100-2, and TNG300-1) show the same results. However, TNG100-1 (the simulation with the higher mass and spatial resolutions), shows qualitatively similar, yet quantitatively different results. The enhancements reported in TNG100-1 are smaller than those of the lower resolution simulations. We note that the dependence on resolution is inconsistent with \cite{TNGpairs} who showed a stronger enhancement in TNG100-1 compared to the lower resolution simulations which is possibly due to the differences in sample selection, control matching, and the SFR metric (SFR within the stellar half mass radius in \citet{TNGpairs}) . Understanding the drivers of the differences between the two simulations (i.e., resolution levels) is beyond the scope of this work. However, we note that \cite{2019MNRAS.485.4817D} report convergence between the two resolution levels (i.e., TNG100-1 and TNG300-1) when using SFRs averaged over timescales $>50$ Myr for galaxies in the same mass range as the post-mergers studied here.


\subsection{The evolution from the pair to post-merger phase}
\noindent
A particular strength of numerical simulations of galaxy mergers is the ability to track the evolution of galaxy mergers and quantify the strength and temporal extent of SFR enhancement and their dependence on galaxy properties. On the contrary, observational studies are unable to accurately follow the evolution of galaxy mergers beyond coalescence; therefore observational studies of post-mergers group the merger remnants together disregarding their evolution\footnote{Observational studies of post-mergers depend on morphological selection which are most sensitive to short timescales beyond coalescence \citep[e.g.,][]{2008MNRAS.391.1137L, 2010MNRAS.404..575L, 2010MNRAS.404..590L}} \citep[e.g.,][]{2015MNRAS.448..221E, 2019MNRAS.482L..55T}. On the other hand, observational studies use the projected galaxy separation as a proxy for time evolution during the interaction/pair phase of galaxy mergers \citep[e.g.,][]{2013MNRAS.435.3627E, 2016MNRAS.461.2589P}. 

\citet{TNGpairs} studied galaxy pairs in Illustris-TNG applying a methodology akin to observations, and reported an enhancement in SFR for close pairs which is in qualitative agreement with observational galaxy pairs studies using SDSS. Knowing that our control matching strategy is different than that used in \citet{TNGpairs}, we reproduce the results of \citet{TNGpairs} following the control matching methods used in this work (i.e., ignoring galaxies with unresolved SFRs, matching within the same class, limiting $r_\mathrm{sep}>2$). Figure \ref{fig:Q_vs_r1-et-Tmerge} combines the results of \citet{TNGpairs} with our results to form a comprehensive picture of galaxy mergers and their evolution in Illustris-TNG. As the separation of galaxy pairs ($r_1$) decreases (pericentric passage), the galaxies experience an enhancement in their SFRs. The enhancement then decays as the galaxies move to their respective apocenters. Following coalescence, the SFR enhancement decays to be consistent with the controls within $\sim 500$ Myr in TNG300-1. We note that the results of the pair phase are possibly underestimated (overestimated) at small (large) separations because, unlike the time domain, at a given separation ($r_1$) we would include the contribution from galaxies which haven not (have) undergone their first pericentric passage. 

%% file: conclusions.tex
\section{Conclusions}
\label{sec:conclusions}
\noindent
We present a large sample of post-mergers simulated in a cosmological environment (selected from the IllustrisTNG simulation suite) and analysed in an observationally motivated scheme (including the use of statistical controls) to demonstrate the effects of galaxy mergers on star formation. We quantify the strength and temporal extent of merger-driven SFR enhancements. Additionally, we tie the enhancement to various galactic properties and hence physical mechanisms. The large sample of post-mergers presented here provides a powerful tool to investigate the detailed evolution of galaxy mergers.  We demonstrated that, for TNG300-1: 
\begin{enumerate}
    \item[$\bullet$]{\textbf{SFR enhancement:} The SFRs of post-merger galaxies, shortly following the merger (i.e., at the first snapshot following the merger), are enhanced by, on average, a factor of $\sim 2$ compared to their controls. The star-forming post-mergers dominate this enhancement, with a factor of $\sim 2$ higher SFRs than their controls. Passive post-mergers exhibit statistically consistent SFRs compared to their controls. }

    \item[$\bullet$]{\textbf{The redshift evolution of merger driven SFR enhancements:} Our post-merger sample exhibits a robust enhancement in SFR (factor of $\sim 2$) for $0 \le z \le 1$, with no evident evolution with redshift. The global enhancement is driven by the star-forming post-merger galaxies, while the passive post-mergers have SFRs which are consistent with those of their controls.}

    \item[$\bullet$]{\textbf{The effect of the mass ratio:} The merger-driven SFR enhancement during the post-merger phase, on average, mildly depends on the mass ratio of the parent galaxies' merger. Minor mergers ($0.1 \le \mu < 0.3$) drive modest enhancements (factor of $\sim 2$) while major mergers ($\mu \ge 0.3$) induce stronger SFR enhancements (factor of $\sim 2.5$).}
    
    \item[$\bullet$]{\textbf{The contribution of minor mergers to the merger-induced SFR budget:} Although major mergers ($\mu \ge 0.3$) drive stronger SFR bursts, they are vastly outnumbered by minor mergers ($0.1 \le \mu <0.3$) which, in spite of their weaker SFR enhancement, still contribute $\sim 50\%$ of the total merger-driven SFR enhancement in TNG300-1.}
    
    \item[$\bullet$]{\textbf{The dependence on stellar mass:} The SFR enhancements in post-merger galaxies show a strong dependence on the hosts' stellar masses. Post-mergers with stellar mass $\log (M_\star / \mathrm{M_\odot}) \le 11.5$ exhibit SFR enhancements of a factor of $\sim 2$ with declining enhancements for larger stellar masses. For post-mergers with stellar masses above $\log (M_\star / \mathrm{M_\odot}) > 11.4$, the dependence on SFR enhancement disappears. The dependence of SFR enhancement on stellar mass is driven by the star-forming post-merger sample; passive post-mergers have SFRs which are, on average, consistent with the SFRs of the controls.}

    \item[$\bullet$]{\textbf{The effects of the merger gas content:} Merger induced star-formation in our post-merger sample (both star-forming and passive) correlates strongly with the merger gas fraction. The descendants of the mergers with the highest gas fractions exhibit the strongest enhancements ($\sim 1.75-2.5$) while the descendants of the lowest gas fraction mergers show suppression in SFR when compared to their controls. }

    \item[$\bullet$]{\textbf{The effect of controlling for gas fraction:} The SFR enhancements in star-forming post-mergers are stronger than their passive counterparts even at the same gas fraction. Moreover, even when constraining the post-merger gas fraction in the control-matching process, SFR enhancements persist, albeit to a weaker extent. The discrepancy in SFR enhancement between the star-forming and passive post-mergers, and the persistence of the SFR enhancement when controlling for the post-merger gas content, are suggestive of additional processes being at play in driving the SFR enhancement (e.g., gas phase, feedback, star-formation efficiency).}

    \item[$\bullet$]{\textbf{The evolution of post-mergers beyond coalescence:} The SFR enhancements in post-mergers decay on a timescale of $\sim 0.5$ Gyr. While the strength of the merger-driven SFR enhancement within $\sim 100-250$ Myr post-coalescence is dependent on the merger mass ratio, the decay in SFR enhancement is \textit{independent of mass ratio} beyond this timescale.}

    \item[$\bullet$]{\textbf{The role of galaxy mergers in suppressing star formation:} Although galaxy mergers do not globally suppress star formation (i.e., quenching), the strongest merger-driven starburst galaxies evolve to be passive/quenched on faster timescales than their controls.}
    
    \item[$\bullet$]{\textbf{Effects of simulation resolution:} The results reported here are qualitatively consistent for the different IllustrisTNG simulation boxes (i.e., resolutions).}
    
\end{enumerate}

%% file: acknowledgements.tex
\section*{Acknowledgments}
The authors thank Connor Bottrell, Jorge Moreno, and Joanna Woo for their insightful comments and helpful discussions. MHH acknowledges the receipt of a Vanier Canada Graduate Scholarship. SLE and DRP acknowledges the receipt of an NSERC Discovery Grant. HG acknowledges Mitacs Globalink. This research was enabled, in part, by the computing resources provided by WestGrid and Compute Canada.